\documentclass[hyper]{JHEP}
\pdfoutput=1
\usepackage{graphicx,amssymb,bm,latexsym,amsmath}
\usepackage{epstopdf}
\usepackage{subcaption}
\usepackage[toc,page]{appendix}
\newcommand{\bej}[1]{ \begin{equation}\label{#1} }
\newcommand{\eej}{\end{equation}}
\newcommand{\beaj}[1]{\begin{eqnarray}\label{#1} }
\newcommand{\eeaj}{\end{eqnarray}}
\newcommand{\eq}[1]{(\ref{#1})}
\def\ZZZ{{\hskip-3pt\hbox{ Z\kern-1.6mm Z}}}
\def\zzz{{\hskip-3pt\hbox{ z\kern-1mm z}}}

\newcommand{\eps}{\epsilon}

\newcommand{\bd}{\bar{\rm D}}

\newcommand{\N}{\frac{m_{2}}{k_{2}}-\frac{m_{1}}{k_{1}}}

\newcommand{\be}{\begin{equation}}
\newcommand{\ee}{\end{equation}}
\newcommand{\ben}{\begin{eqnarray}\displaystyle}
\newcommand{\een}{\end{eqnarray}}

\def\one{{\hbox{ 1\kern-.8mm l}}}
\def\zero{{\hbox{ 0\kern-1.5mm 0}}}

\def\be{\begin{equation}}       
\def\ee{\end{equation}}         

\def\bea{\begin{eqnarray}}      
\def\eea{\end{eqnarray}}
                                
\def\ba{\begin{array}}
\def\ea{\end{array}}
\def\bd{\begin{displaymath}}
\def\ed{\end{displaymath}}

\def\eq{\begin{equation}}
\def\eqe{\end{equation}}
\def\eqa{\begin{eqnarray}}
\def\eqae{\end{eqnarray}}
\def\ena{\end{eqnarray}}

\def\unit{1 \hskip-.3em \raise2pt\hbox{$ \scriptstyle |$ } }

\def\s{\sigma}                                   

\def\bd{\begin{displaymath}}
\def\ed{\end{displaymath}}

\def\6{\partial}
\def\N4{{\cal N}=4}

\def\bop#1{\setbox0=\hbox{$#1M$}\mkern1.5mu
        \vbox{\hrule height0pt depth.04\ht0
        \hbox{\vrule width.04\ht0 height.9\ht0 \kern.9\ht0
        \vrule width.04\ht0}\hrule height.04\ht0}\mkern1.5mu}

\def\>{\rangle} 

\def\<{\langle} 
\def\Dsl{D \hskip-.6em \raise1pt\hbox{$ / $ } }
\def\to{\rightarrow}

\def\+{\oplus}

\def\as2{AdS_3\times S^3_1 \times S^3_2}

\title{\begin{flushright}
\normalsize {YITP-18-69}\\
\end{flushright}Probing analytical and numerical integrability: The curious case of $(AdS_5\times S^5)_{\eta}$}

\author{
{\bf {\normalsize Aritra Banerjee,$^{a}$ Arpan Bhattacharyya,$^{b}$ }
\thanks{E-mail: aritra@itp.ac.cn, bhattacharyya.arpan@yahoo.com}
}
\\
{\normalsize $^{a}$CAS Key Laboratory of Theoretical Physics,}\\{\normalsize Institute of Theoretical Physics,}\\{\normalsize Chinese Academy of Sciences,  Beijing 100190, P.R.China}
\\
 {\normalsize $^{b}$ Center for Gravitational Physics, Yukawa Institute for Theoretical }\\{\normalsize Physics,
Kyoto University, Kyoto 606-8502, Japan.}
 
}

\abstract{Motivated by recent studies related to integrability of string motion in various backgrounds via analytical and numerical procedures, we discuss these procedures for a well known integrable string background $(AdS_5\times S^5)_{\eta}$. We start by revisiting conclusions from earlier studies on string motion in $(\mathbb{R}\times S^3)_{\eta}$ and $(AdS_3)_{\eta}$ and then move on to more complex problems of  $(\mathbb{R}\times S^5)_{\eta}$ and $(AdS_5)_{\eta}$. Discussing both analytically and numerically, we deduce that while $(AdS_5)_{\eta}$ strings do not  encounter any irregular trajectories, string motion in the deformed five-sphere can indeed, quite surprisingly, run into chaotic trajectories. We discuss the implications of these results both on the procedures used and the background itself. }


\begin{document}
\section{Introduction}
String motion in curved spaces, described by two-dimensional non-linear sigma models, have been studied extensively from the early days of development of the subject. This is extremely interesting due to the complicated non-linear equations of motion associated with the worldsheet fields. It comes as no surprise that these equations of motion are only `integrable' for a select subclass of target space backgrounds, and hence this notion of integrability helps one to pick out the cases where a complete quantitative analysis of classical (and perhaps quantum) string motion can be performed and compared to the flat space case. One of the most widely known cases is of course that of type IIB strings in the $AdS_5\times S^5$ space-time \cite{Bena:2003wd},  which is dual to operators in maximally supersymmetric $\mathcal{N} = 4$ Yang-Mills theory (sYM) via  $AdS/CFT$ correspondence \cite{Maldacena:1997re}. The integrability of these strings moving in the bulk $AdS_5\times S^5$, in conjunction with the integrability of the dual sYM theory, makes an exceptional example to study the $AdS/CFT$ correspondence from  the point of view  of integrable systems \cite{Beisert:2010jr}. Moreover, the finding that in the $semiclassical$ limit, the dynamics of this correspondence indeed becomes tractable \cite{Gubser1}, has regenerated interest in the classical string solutions in $AdS$ and related geometries. Indeed, a lot of literature has been devoted to the subject of integrability in $AdS/CFT$ in last two decades\footnote{For recent introductions
to this subject, the reader is directed to \cite{vanTongeren1, Bombardelli:2016rwb} and references therein.}.  

With this advent of integrability studies in the context of $AdS/CFT$, there have been many celebrated works relating to deformation of the symmetries on both sides of the correspondence while keeping the integrable structure intact. Most of these relied on the use of target space duality symmetries to generate new integrable backgrounds \cite{Lunin:2005jy,Frolov:2005ty,Frolov:2005dj,Alday:2005ww}. Recently,  Klimcik's pioneering works on novel integrable deformations of $\sigma$-models \cite{Klimcik:2002zj, Klimcik:2008eq,Klimcik:2014bta} have paved the way for their application to string $\sigma$-models and finding probable deformed versions of $AdS/CFT$ correspondence \cite{Delduc1}. Since then  a larger family of integrable deformations of $AdS\times S$ geometries have been explored, where the deformation is given by a classical $r$-matrix solution to the (modified) Classical Yang-Baxter Equation (CYBE). The explicit geometry and NS-NS forms for such a `Yang-Baxter' deformation of $AdS_5\times S^5$ first appeared in \cite{Delduc1, Aru1}, was analysed in detailed in \cite{Delduc2} and various consistent truncations have been discussed in \cite{Hoare1}. In the Yang-Baxter case, the deformation works by deforming the supercoset associated to $AdS_5 \times S^5$ itself by a continuous parameter, which is often referred to as a $q$-deformation, or a quantum group deformation \cite{Delduc2}. This replaces the lie algebra of the classical charges by its $q$-deformed version, which is then incorporated into the superstring action for $AdS_5 \times S^5$ having a real deformation parameter $\eta \in [0,1) $ or equivalently another parameter called $\kappa$ with
$\kappa \in [0,\infty)$. Notice that here the parameter $ \kappa $ is related to the original deformation parameter $ \eta $ as 
$\kappa = \frac{2\eta}{1-\eta^{2}}$ \cite{Aru1}. In the rest of the paper, we would instead denote $ \kappa $ as being the deformation parameter in our analysis and refer the background as  $(AdS_5\times S^5)_\kappa$. For various avenues of exploratory works on Yang-Baxter deformations, one should have a look at \cite{Kawaguchi1}-\cite{Crichigno1}.

As integrable string backgrounds by construction, the conserved currents associated to $(AdS_5 \times S^5)_\kappa$ strings, in general, satisfy the Lax equations. However in the case of a random string sigma model, where the existence of Lax pair is not known, proving (non)-integrability is a rather complicated task. To this note, there have been a number of works to consistently truncate the two-dimensional string equations of motion of particular circular strings into one dimensional mechanical systems and analyzing the (non)-integrability properties thereof. It has been argued that it is sufficient to show that there exists at least one truncated dynamical system of differential equations, where the corresponding string motion turns $chaotic$ \cite{Basu:2011di}, i.e. small variations around the equations grow non-deterministically in time. Useful tools in these studies have mainly been the variational non-integrability techniques of Hamiltonian systems and numerical experiments in the associated phase space in general. This approach is often hailed as the equivalent of the algebraic approach of finding Lax pairs for the system and a large number of works have appeared along these lines, see for example \cite{Zayas:2010fs}-\cite{Nunez:2018qcj}.  

In the following note, we seek to understand this equivalence by studying string motion in the extremely complicated but integrable background of $(AdS_5 \times S^5)_\kappa$. One should bear in mind, the process of Yang-Baxter deformation breaks the supersymmetries associated with $(AdS_5 \times S^5)$ and even at the bosonic level, the isometry group of $SO(2,4)\times SO(6)$ breaks down to $U(1)^3\times U(1)^3$ in this case, but still the $\kappa$-deformed background inherits the parent integrability. We must mention here \cite{Roychowdhury:2017vdo}, in which, using the above-mentioned hamiltonian analytical methods, it was claimed that the associated phase space encounters chaos as the differential equations of motion are not `integrable'. This certainly creates a tension between the different methods of studying (non-)integrability of string motion in curved backgrounds. Spearheaded by this, we revisit these claims of non-integrability of string motion in  $(\mathbb{R}\times S^3)_{\kappa}$ and $(AdS_3)_{\kappa}$ and then attack the larger and more complicated problem of strings in $(\mathbb{R}\times S^5)_{\kappa}$ and $(AdS_5)_{\kappa}$ with antisymmetric $B$ fields included. We explicitly show that string motion in the former cases $does~not$ have any non-integrable traits. However, to our surprise, we find that the phase space of the deformed five-sphere indeed contains chaotic string motion, as the equations describing the motion are non-integrable in nature. We emphasize that this phenomenon happens only for the dynamical phase space associated with the full five sphere, and the sub-sectors can be presumed integrable in this sense. For the case of  $(AdS_5)_{\kappa}$ also, we find string trajectories remain regular throughout the motion.

The paper is organized in the following way, in  section 2, we give a review of the background and fluxes associated to the $(AdS_5 \times S^5)_\kappa$ string background. In section 3, after revisiting the case of $(\mathbb{R}\times S^3)_{\kappa}$,  we will have a detailed discussion of  string motion in $(\mathbb{R}\times S^5)_{\kappa}$. By the use of Normal Variational Equations (NVE's) for fluctuations around equations of motion for $consistent$ string solutions, we would arrive at the fact that strings in $(\mathbb{R}\times S^3)_{\kappa}$ do not run into any chaotic trajectories. In the case of deformed five-sphere, we will, however, find chaotic trajectories as soon as we turn on a non-zero deformation parameter, a result that will be corroborated by both using NVE and studying its Poincare sections by numerical trajectories method. In section 4, we will essentially repeat the same exercise of the earlier chapter but for the case of deformed $AdS$ backgrounds. As in the earlier case, we show that there are no chaotic trajectories in $(AdS_3)_{\kappa}$. And following this, no chaotic motion is found in the case of $(AdS_5)_{\kappa}$ as well, which we confirm via both analytical and numerical calculations. We discuss the ramifications of our results and conclude this work in section 5.
\section{Setup}
Let us start by introducing the geometry and the general setup required for our study.
We first write down the full deformed metric for the $\kappa$ deformed $AdS_5\times S^5$ \cite{Aru1},

\begin{eqnarray}\label{eq:metrc-etaAdS5S5-sph-coord}
\begin{aligned}
	ds^2_{(\text{AdS}_5)_{\kappa}}=&-\frac{1+\rho^2}{1-\kappa^2\rho^2}dt^2
	+\frac{d\rho^2}{ \left(1+\rho^2\right)(1-\kappa^2\rho^2)}\\
	&  + \frac{\rho^2}{1+\kappa^2\rho^4\sin^2\zeta}\left( d\zeta^2+\cos ^2\zeta \, d\psi_1^2\right) 
	+\rho^2 \sin^2\zeta\, d\psi_2^2\,,
	\\
	\\
	ds^2_{(\text{S}^5)_{\kappa}}=&\frac{1-r^2}{1+\kappa^2 r^2}d\phi^2
	+\frac{dr^2}{ \left(1-r^2\right)(1+\kappa^2 r^2)}\\
	&  + \frac{r^2}{1+\kappa^2r^4\sin^2\xi}\left( d\xi^2+\cos ^2\xi \, d\phi_1^2\right) 
	+r^2 \sin^2\xi\, d\phi_2^2\,.
\end{aligned}
\end{eqnarray}
Also we have the $B$-fields $B=\frac{1}{2} B_{MN}\ dX^M\wedge dX^N$~\cite{Aru1} associated to the solution,
\begin{eqnarray}\label{eq:B-field-etaAdS5S5-sph-coord}
\begin{aligned}
	\widetilde{B}_{(\text{AdS}_5)_{\kappa}} &= +\frac{\kappa}{2} \left( \frac{\rho^4 \sin (2\zeta)}{1+\kappa^2 \rho^4\sin^2 \zeta} d\psi_1\wedge d\zeta + \frac{2 \rho}{1-\kappa^2 \rho^2}dt\wedge d\rho\right),
	\\
	\widetilde{B}_{(\text{S}_5)_{\kappa}} &= -\frac{\kappa}{2} \left( \frac{r^4 \sin (2\xi)}{1+\kappa^2 r^4\sin^2 \xi}d\phi_1\wedge d\xi + \frac{2r}{1+\kappa^2 r^2}d\phi\wedge dr\right).
\end{aligned}
\end{eqnarray}
It is easy to see that the contributions of the components $B_{t\rho}$ and $B_{\phi r}$ to the Lagrangian are total derivatives, and hence can be ignored without loss of generality.
We can put in $\rho = 0$ and $r= \cos\theta$ and perform the redefinition of the coordinates $\phi \to \phi_3$ and $\xi \to \psi$ to write the metric of $(\mathbb{R}\times S^5)_{\kappa}$ in the following form, 
 \begin{eqnarray}\label{newmetric}
\begin{aligned}
	\\
	ds^2_{(\mathbb{R}\times S^5)_{\kappa}}=&-dt^2+\frac{\sin^2\theta}{1+\kappa^2 \cos^2\theta}d\phi_3^2
	+\frac{d \theta^2}{1+\kappa^2 \cos^2\theta}\\
	&  + \frac{\cos^2\theta}{1+\kappa^2 \cos^4\theta\sin^2\psi}\left( d\psi^2+\cos ^2\psi \, d\phi_1^2\right) 
	+\cos^2\theta \sin^2\psi\, d\phi_2^2\,.
\end{aligned}
\end{eqnarray}
And in this case, the single surviving component of NS-NS flux takes the form as,
\begin{eqnarray}\label{Bfield}
\begin{aligned}
	\\
	\widetilde{B}_{(\mathbb{R}\times S^5)_{\kappa}} &= -\frac{\kappa}{2} \left( \frac{\cos^4\theta \sin (2\psi)}{1+\kappa^2 \cos^4\theta\sin^2 \psi}d\phi_1\wedge d\psi \right).
\end{aligned}
\end{eqnarray}

It is worthwhile to note that the $(AdS)_{\eta}$ contains a singularity, but we won't be bothered with that part in the present analysis. We write the deformed $(AdS_5)_\kappa$ part of the metric and $B$ field again with the redefinition $\rho \to \sinh\rho$, 
\bea\label{newmetric2}
ds^2_{(AdS_5)_\kappa} &=&\frac{1}{1-\kappa^2\sinh^2\rho}\left[ -\cosh^2\rho dt^2 + d\rho^2 \right] \nonumber
	\\
	&+& \frac{\sinh^2 \rho}{1+\kappa^2\sinh^4 \rho \sin^2\zeta}\left( d\zeta^2+\cos ^2\zeta \, d\psi_1^2\right) 
	+\sinh^2\rho \sin^2\zeta\, d\psi_2^2\,\nonumber
	\\
B &=& +\frac{\kappa}{2} \left( \frac{\sinh^4\rho \sin (2\zeta)}{1+\kappa^2 \sinh^4\rho \sin^2 \zeta} d\psi_1\wedge d\zeta \right).
\eea
The singularity surface in this coordinate system is located at a critical value of the radial coordinate
\be
\rho = \rho_s = \sinh^{-1}\frac{1}{\kappa},
\ee
So that $\kappa \to 0$ signals the usual $AdS$ boundary at conformal infinity. One must emphasize, that this is a general singularity in the spacetime which cannot be dealt with by simple change of coordinates alone.

In this background, to study string solutions, we use the Polyakov action coupled to an antisymmetric B-field, 
\begin{eqnarray} \label{lag1}
S &=&\int d\sigma d\tau~(\mathcal{L}_G +\mathcal{L}_B)\\ \nonumber
&=&-\frac{\sqrt{\hat\lambda}}{4\pi}\int d\sigma d\tau
[\sqrt{-\gamma}\gamma^{\alpha \beta}g_{MN}\partial_{\alpha} X^M
\partial_{\beta}X^N - \epsilon^{\alpha \beta}\partial_{\alpha} X^M
\partial_{\beta}X^N B_{MN}] \ ,
\end{eqnarray}
where  $\sqrt{\hat\lambda}$ is the modified 't Hooft coupling for this case, given by $\hat\lambda = \lambda(1+\kappa^2)^{1/2}$, \footnote{ The radius of deformed $AdS$ and undeformed one are related as $R_{\kappa}^2 = R_{AdS}^2(1+\kappa^2)^{1/2}$. In a case analogous to $AdS/CFT$, one could then relate the radius of the deformed sphere (or $AdS$) to a `modified' `t Hooft coupling. However, no such concrete proof of a dual gauge theory for this background has been found yet.}
$\gamma^{\alpha \beta}$ is the worldsheet metric and $\epsilon^{\alpha
	\beta}$ is the antisymmetric tensor defined as $\epsilon^{\tau
	\sigma}=-\epsilon^{\sigma \tau}=1$. $X^M(\tau,\sigma)$ are bosonic worldsheet fields and they are functions of the two worldsheet coordinates $\tau$ and $\sigma.$ Variation of the action with respect to these  $X^M(\tau,\sigma)$ gives us the following equations of motion,
\begin{eqnarray}
2\partial_{\alpha}(\eta^{\alpha \beta} \partial_{\beta}X^Ng_{KN})
&-& \eta^{\alpha \beta} \partial_{\alpha} X^M \partial_{\beta}
X^N\partial_K g_{MN} - 2\partial_{\alpha}(\epsilon^{\alpha \beta}
\partial_{\beta}X^N b_{KN}) \nonumber \\ &+& \epsilon ^{\alpha \beta}
\partial_{\alpha} X^M \partial_{\beta} X^N\partial_K b_{MN}=0 \ ,
\end{eqnarray}
and variation with respect to the metric gives the two Virasoro
constraints,
\begin{eqnarray}
g_{MN}(\partial_{\tau}X^M \partial_{\tau}X^N +
\partial_{\sigma}X^M \partial_{\sigma}X^N)&=&0 \label{v1} \\ 
g_{MN}(\partial_{\tau}X^M \partial_{\sigma}X^N)&=&0 \label{v2}\ .
\end{eqnarray} \label{VirCons}
Where the first constraint is equivalent to the vanishing of the two-dimensional Hamiltonian density. We use the conformal gauge (i.e.
$\sqrt{-\gamma}\gamma^{\alpha \beta}=\eta^{\alpha \beta}$) with
$\eta^{\tau \tau}=-1$, $\eta^{\sigma \sigma}=1$ and $\eta^{\tau
\sigma}=\eta^{\sigma \tau}=0$) to solve these equations of motion.	For the case of  $(AdS_5)_{\kappa} \times (S^5)_{\kappa}$ we identify, $X^M$ for $M=1,\cdots 5$ with the coordinates of  $(AdS_5)_{\kappa}$ i.e with $t, \rho,\zeta,\psi_1,\psi_2$ and $X^{M}$ for $M=6,\cdots 10$ with coordinates of $(S^5)_{\kappa}$ i.e with $\phi,r,\xi,\phi_1,\phi_2.$ Hence the components of $g_{MN}$ with $M, N$ from $1,\cdots 5$ are the components of the $(AdS_5)_{\kappa}$ metric and $g_{MN}$ with $M,N$ form $6,\cdots 10$ are the components of $(S^{5})_{\kappa}$ metric. Similar identifications can be made for all other cases. 
\section{Strings in deformed sphere}

\subsection{Revisiting a warm-up example: The case of $(\mathbb{R}\times S^3)_{\kappa}$}

Although the simplest case of an extended string in  $(\mathbb{R}\times S^3)_{\kappa}$ has been addressed already in \cite{Roychowdhury:2017vdo}, we would first start with taking another look at the findings. We start with the following metric,
\be
ds^2_{(\mathbb{R}\times S^3)_{\kappa}} = -dt^2
	 + \frac{1}{1+\kappa^2 \cos^2\theta}\left( d\theta^2+\sin^2\theta\, d\varphi^2\right) +\cos^2\theta\, d\phi^2.
\ee

The NS-NS flux vanishes in this case. We now have to choose a consistent worldsheet embedding for the worldsheet coordinates. We must note here that in \cite{Roychowdhury:2017vdo} the following string embedding was chosen,
\be \label{prof}
t=t(\tau),~~\theta = \theta(\tau),~~\varphi(\tau,\sigma) =  \alpha_1\sigma + q(\tau),~~\phi = \alpha_2 \sigma.
\ee
We must mention here that this does not seem to be a consistent string embedding, since for this choice, the second Virasoro constraint ($T_{\tau\sigma}= 0$) gives rise to the following condition,
\be
\frac{\sin^2\theta}{1+\kappa^2 \cos^2\theta}~\alpha_1\dot{q} = 0,
\ee
Which in turn can  only be satisfied consistently if the winding number $\alpha_1 = 0$ or $\dot{q} = 0$. For our case, we choose the former  and propose a refined ansatz for a circular string with additional angular momentum in the aforementioned geometry,
\be
t=t(\tau),~~\theta = \theta(\tau),~~\varphi =  \varphi(\tau),~~\phi = m \sigma.
\ee
 This is a completely consistent  embedding and makes the second Virasoro constraint zero naturally. This is one of the simplest possible rigid string embeddings in the sense that just by putting the winding number along the direction $\phi$ to be zero ($m=0$), we could have the action of a simple particle moving in this spacetime. This simple consideration guarantees that the worldsheet theory can be truncated to one dimensional dynamical system consisting of a system of non-linear ODE's.  In general, one can expect if chaotic behaviour shows up in the simplest string configuration, it will suffice to show non-integrability of all string motion in the background. We can now write 
the effective Lagrangian of this theory as,
\bea
\mathcal{L}_S  = -\dot{t}^2 + \frac{\dot\theta^2}{1+\kappa^2 \cos^2\theta}+ \frac{\sin^2\theta~\dot{\varphi}^2}{1+\kappa^2 \cos^2\theta}-m^2 \cos^2\theta.
\eea
From the equations of motion, it can be seen that the $t$ equation is easily satisfied by,
\be
t = E\tau,
\ee
where $E$ is a constant, and the $\phi$ equation is trivially satisfied. The other two equations for $\theta$ and $\varphi$ then read as following,
\bea
-\sin\theta \cos\theta \left[m^2\left(1+\kappa ^2  \cos ^2\theta\right)^2-\kappa ^2 \dot\theta^2+\left(1+\kappa ^2\right) \dot\varphi^2\right]+ \ddot\theta \left(1+\kappa ^2 \cos ^2\theta\right) &=& 0\,\nonumber \\ 
2\dot\theta  \dot\varphi \sin \theta\cos\theta \left(1+\kappa ^2\right) +\ddot\varphi \sin^2\theta \left(1+\kappa^2\cos^2\theta\right)&=& 0. \,\nonumber
\eea
These equations are supplemented by the other Virasoro constraint, which can be written as, 
\be
E^2 = \frac{\dot\theta^2}{1+\kappa^2 \cos^2\theta}+ \frac{\sin^2\theta~\dot{\varphi}^2}{1+\kappa^2 \cos^2\theta}+m^2 \cos^2\theta.
\ee
One can actually check that the time derivative of the Virasoro constraint vanishes when it is evaluated using the second order equations of motion, making our ansatz consistent with both equations of motion and Virasoro constraints. From the above, we can see that $\theta\to0,~\dot\theta\to0$ is a solution to the both equations of motion, i.e. it defines an invariant plane of the system. We can demand that the Hamiltonian constraint is satisfied on the invariant plane, just with the identification of the constants  as $E^2 = m^2$. Now we can consider small fluctuations around this invariant plane, with the form,
\be
\theta(\tau) = 0 +\eps(\tau),~~~|\eps|<<1.
\ee
Expanding the $\theta$ equation upto first order in $\eps$, we can get the Normal Variational Equation(NVE) for $\theta$,
\be \label{NVE1}
\ddot{\eps}-\left(m^2(1+\kappa^2)  +\dot\varphi^2  \right)\eps=0.
\ee
The question of integrability of the string motion has now turned into whether or
not the above NVE can be solved in quadratures. We can consider a special class of solutions of the NVE which in turn are functions of exponentials, logarithms and/or algebraic expressions (or integrals of such functions) of the independent variables of the system. These classes of solutions are often called $Liouvillian$ solutions \cite{Stepanchuk:2012xi}. A NVE that does not admit Liouvillian solutions indicates that the underlying dynamical system is non-integrable. 

We now have to replace $\dot{\varphi}$ to get a differential equation for $\eps(\tau)$.   We then note that from the equation for $\varphi$ we can easily write,
\be
\partial_{\tau}\Big[ \frac{\sin^2\theta~\dot{\varphi}}{1+\kappa^2\cos^2\theta}\Big] = 0 \rightarrow \dot\varphi = J \Big[ \frac{1+\kappa^2\cos^2\theta}{\sin^2\theta}\Big]
\ee
where $J$ is a constant of motion, i.e. evolves independently of time. So, near the point $\theta \to 0$, we can write
\be
\dot\varphi \sim \frac{J(1+\kappa^2)}{\eps^2}.
\ee
This simply suggests that $J\to 0$ as we go near the $\theta \to 0$ plane. With this replacement, we can now analyze the NVE to find that it offers well defined Liouvillian solutions of the form,
\be
\eps(\tau)= \pm~\frac{\sqrt{4 m^2 \left(1+\kappa ^2\right)^3 J^2+e^{-4 m \sqrt{1+\kappa ^2} \tau}}}{2 m \sqrt{1+\kappa ^2} \sqrt{e^{-2 m \sqrt{1+\kappa ^2} \tau}}}.
\ee
The above solution is completely well defined in the parameter space and we can conclude that the string motion, in this case, does not run into chaos anywhere. Note that if we had chosen $\theta =\pi/2$ as an invariant plane of the system, we would have simply gotten $\dot\varphi \sim J$, which would also be sufficient to satisfy the equations of motion and the NVE would just take the form of a simple Harmonic Oscillator equation. The Hamiltonian constraint, in that case, would simply become $E^2 = J^2$, i.e that of a BPS point-like string. In the above analysis, we focussed on the more non-trivial case for having a better picture. 

For the sake of completeness let us also discuss the case of choosing the angular momentum along the other isometry direction from the above one, i.e considering the changed ansatz, 
\be
t=E\,\tau ,~~\theta = \theta(\tau),~~\varphi = m \sigma ,~~\phi = \varphi(\tau),
\ee
since the two-spheres inside the deformed three-sphere are not equivalent to each other (as is the case  for undeformed spheres) this case has to be addressed separately. In this case, the equations of motion can be seen to be trivially satisfied by $\theta \to \frac{\pi}{2},~\dot\theta \to 0$, which define the other invariant plane. Doing the above analysis again for this case (expanding as $\theta(\tau) = \frac{\pi}{2} +\tilde\eps(\tau)$) yields the same form of NVE as in (\ref{NVE1}).
The only difference comes from the definition of the angular momenta $\tilde J$, which near $\theta \to \frac{\pi}{2}$ gives,
\be
\dot\phi \sim \frac{\tilde J}{\tilde\eps^2}.
\ee
 So we can safely say here that the expansion near invariant planes is not sensitive to the choice of the angular momentum direction, and in both the cases integrability properties of the equations of motion stay unchanged.
\subsection{Strings in the five-sphere: Analytical}

In this section, we would try to repeat the exercise done in the last section for a deformed $S^5$. Due to the complexity of the equations of motion in this case, more emphasis will be given to the numerical analysis performed in the next subsection. Let us take a general spinning string ansatz \footnote{A similar ansatz was used to study non-integrable trajectories in complex-$\beta$ deformed five-sphere in \cite{Giataganas2013}.} in the deformed five-sphere of the following form,
\be \label{ans1}
t =E\,\tau,~\theta = \theta(\tau),~\psi = \psi(\tau),~\phi_3 =\phi_2=0,~\phi_1 =m\s.
\ee
The effective lagrangian of the theory is given by,
\bea \label{lag1}
\mathcal{L}_S  = -\dot{t}^2 + \frac{\dot\theta^2}{1+\kappa^2 \cos^2\theta}+ \frac{\cos^2\theta~(\dot\psi^2-\cos^2\psi~m^2)}{1+\kappa^2 \cos^4\theta\sin^2\psi}+ \frac{\kappa\cos^4\theta \sin (2\psi)~m\dot\psi}{1+\kappa^2 \cos^4\theta\sin^2 \psi}.
\eea

The equation of motion for $t$ is satisfied trivially. The equation of motion for $\theta$ on the other hand reads,
\bea\label{theq}
\frac{\sin \theta \cos \theta \left(\dot\psi^2-m^2 \cos ^2\psi\right)}{1+\kappa ^2 \cos ^4\theta \sin ^2\psi}+\frac{2 \kappa ^2 \sin \theta  \cos ^5\theta  \sin ^2\psi \left(m^2 \cos ^2\psi-\dot\psi^2\right)}{\left(1+\kappa ^2 \cos ^4\theta \sin ^2\psi \right)^2}&+&\\ \nonumber 
\frac{2 \kappa  m \sin\theta \cos^3\theta~ \dot\psi \sin (2 \psi)}{1+\kappa ^2 \cos ^4\theta \sin ^2\psi }-\frac{4 \kappa ^3 m \sin \theta\cos ^7\theta  ~\dot\psi \sin ^3\psi  \cos \psi}{\left(1+\kappa ^2 \cos ^4\theta  \sin ^2\psi\right)^2}&+& \\ \nonumber
\frac{\ddot\theta}{1+\kappa ^2 \cos ^2\theta}+\frac{\kappa ^2 \dot\theta^2 \sin \theta \cos\theta}{\left(1+\kappa ^2 \cos ^2\theta\right)^2}&=& 0.
\eea
Similarly, we can easily write the equation of motion for $\psi$ as,
\bea
\cos^2 \theta \Big[\sin (2 \psi) \left(-m^2 \left(3 \kappa ^2+\kappa ^2 (4 \cos (2 \theta)+\cos (4 \theta))+8\right)-8 \kappa ^2 \cos ^4\theta\dot\psi^2\right)&+& \\ \nonumber 
16 ~\ddot\psi \left(1+\kappa ^2 \cos ^4\theta \sin^2\psi\right)-32~ \dot\theta \sin \theta \big(\kappa  m \cos ^2\theta \sin (2 \psi)&+& \\  \nonumber
 \dot\psi \left(1-\kappa ^2 \cos ^4\theta \sin ^2\psi\right)\big)\Big] &=& 0. \nonumber
\eea
In these equations, we can see that $\theta = 0$ and $\psi = \frac{\pi}{2}$ are trivial solutions of the $\theta$ and $\psi$ equations respectively. The non-zero Virasoro constraint equation has a form, 
\be \label{vir2}
E^2 = \frac{\dot\theta^2}{1+\kappa^2 \cos^2\theta}+ \frac{\cos^2\theta~(\dot\psi^2+\cos^2\psi~m^2)}{1+\kappa^2 \cos^4\theta\sin^2\psi}.
\ee
\subsection*{Normal variational equations}
Let us first consider the equation of motion for $\psi$. This is satisfied by the trivial solution as we showed earlier. If we now look carefully at $\psi = \frac{\pi}{2}$ solution, the Virasoro constraint  on the invariant dynamical plane yields,
\be\label{inveq}
\dot\theta^2 = (1+\kappa^2\cos^2\theta)E^2,
\ee
thereby effectively eliminating one variable from the equation. For us,  $\psi = \frac{\pi}{2}$ will be the aforementioned invariant plane, making the dynamics effectively only along the $\theta$ direction. This special solution to the equations of motion is given by (\ref{inveq}), which can be solved with appropriate initial conditions to get,
\be \label{sol1}
\theta(\tau) = \bar\theta=  \mathbf{sn}~\Big[ \sqrt{1+\kappa^2}E\tau~|~\frac{\kappa^2}{1+\kappa^2}\Big].
\ee
Now we have to study small fluctuations near this special solution in order to comment on the integrability of the system. To write the normal variation equation, we then start by expanding the equation of motion for $\psi$ using 
\be
\psi(\tau) = \frac{\pi}{2}+ \eta(\tau),~~~|\eta|<<1.
\ee
Upto first order in $\eta$ the expansion then reads,
\bea
\ddot\eta(1+ \kappa^2 \cos^4\theta) &+& 2(\kappa^2\cos^3\theta\sin\theta-\tan\theta) ~\dot\theta \dot\eta \\ \nonumber  &-&\Big[m^2 \big(1+ \kappa^2 \cos^4\theta) +
2~m\kappa\sin(2\theta)\dot\theta \Big]\eta=0.
\eea
One should note since there is a division by $\cos^2\theta$ involved, this expansion does not include the case of $\theta =\frac{\pi}{2}$. Also since we are working in strictly first order of $\eta$ in this paper, we can discard terms containing $\eta^2,~\dot\eta^2$ etc. 
Now, on the invariant plane, we can  simply use (\ref{inveq}) to replace terms having derivatives of $\theta$
Instead of working with the total complicated differential equation, we can concentrate on the $\kappa\to 0$ limit, i.e. that of small deformation. We would see that this would be enough for our case. We can write this differential equation upto the leading order of $\kappa$ in the following form,
\be
\eta''+ \frac{m^2\eta}{E^2(1+z^2)^2}+\frac{4m\kappa  z\eta}{E(1+z^2)^3}+\mathcal{O}(\kappa^2) = 0,
\ee 
where we have done a change of variables of the form  $\tau \to z= \tan\theta(\tau)$ to express the coefficients of the differential equation in rational form.
This equation is already in the so-called ``Normal'' or Schr\"{o}dinger form.  It is easy to see that the $\kappa=0$ solution, i.e. the solution for the undeformed sphere is quite simple and in a Liouvillian form
\be
\eta^{(0)} =\frac{1}{2} \sqrt{1+z^2}~ e^{-i \sqrt{\frac{m^2}{E^2}+1} \tan ^{-1}(z)} \left(2 c_1 e^{2 i \sqrt{\frac{m^2}{E^2}+1} \tan ^{-1}(z)}+\frac{i c_2}{\sqrt{\frac{m^2}{E^2}+1}}\right)
\ee
Where $c_i$ are constants. But the total solution $\eta^{(\kappa)}$ cannot be written in terms of rational functions. Specifically, the solutions can be found in terms of Doubly Confluent Heun function, thereby making it clear that even in the small deformation limit, there is no Liouvillian solution for this dynamics, making it effectively non-integrable in this sense. This claim will be more elucidated in the next section when we talk about numerical simulation of these string trajectories. 

Of course, the above discussion does not encompass the whole story here. In general, the integrability properties of classical Hamiltonian systems are associated with the behaviour of variations for the phase space curves. The usefulness of NVE's come in handy when we systematically want to analyze the existence of functionally independent integrals of motion. The symmetries leading to existence of such integrals of motion are usually given by transformations between space of solutions of the variational differential equations. These are often described in mathematical literature via Picard-Vessiot theory or differential Galois group techniques \cite{Galois}(Also see \cite{Stepanchuk:2012xi}). Since determining the Galois group for a general case is hard, a different route is provided via the Koavacic algorithm \cite{Kovacic} to find the existence of Liouvillian solutions. We will describe more about this in the appendix, and explicitly calculate the case of $\theta$ NVE in $(\mathbb{R}\times S^5)_{\kappa}$ for any finite value of $\kappa$ via this algorithm, which will undoubtedly point out that inclusion of non-zero $\kappa$ leads to solutions becoming non-Liouvillian in this case. For the time being, we will accept the above discussion and focus on the numerical analysis. 

\subsection{The hamiltonian and numerical trajectories}
Here we supplement our previous analysis by probing more into the chaotic behaviour. We  will find the Hamiltonian equation of motion  and plot the constant energy surfaces.  Then, by observing the behaviour of those trajectories we can get some insight into this chaotic behaviour by invoking the Kolomogorov-Arnold-Moser (KAM) theorem. For integrable systems essentially the number of conserved charges is equal to the number of degrees of  freedom present in the system. The systems that we will consider are basically coupled harmonic oscillators with non-trivial potentials. They are characterized  by a set of coordinates $q_i$ and their conjugate momenta $p_i.$ Together the set of $\{q_i,p_i\}$ give the phase space $(i=1,\cdots N ).$ Now if the system is integrable then there will be exactly $N$ number of conserved charges. Then we can plot this $N$ dimensional surfaces and typically for the integrable system the shape of these surfaces are of like a torus, which is known as KAM tori. In other words for each value of these conserved charges (one of them will be the energy which we will mainly consider in our subsequent analysis), the points of the phase space will lie on this KAM tori. \par
Now when one adds non-integrable terms to the integrable Hamiltonian these KAM tori get perturbed.  According to the KAM theorem most of these tori will be deformed but if the strength of the non-integrable deformation terms is small then the trajectories will still  be ordered and fall on the surface of this deformed tori (only the resonant tori i.e those corresponding to the frequencies $\omega_i$ such that $\alpha^i\omega_i=0$, where $\alpha^i \in Q$ will be completely destroyed). But if the strength of the non-integrable deformations is large, all these tori will be completely destroyed and the trajectories can probe the entire accessible phase space (determined by the total energy)  in a completely arbitrary way and thus we will observe chaotic behaviour.\par
We will adopt the following strategy in our case. We first consider the string motion on $(\mathbb{R}\times S^5)_{\kappa}$ case as discussed in section (3.2) and use the profile mentioned in (\ref{ans1}). We write down the Hamiltonian starting from the Lagrangian mentioned in (\ref{lag1}) below.
\begin{align}
\begin{split}
\mathcal{H}_{S^5_{\kappa}}=\frac{1}{4}  \Big[p_{\theta}^2 \left(1+\kappa^2 \cos^2\theta \right)+\frac{p_{\psi}^2 }{\cos^2\theta}-E^2 +\cos^2\theta(\kappa\,  p_{\psi} \sin \psi-2 m \cos ^2\psi )\Big],
\end{split}
\end{align}
where the two conjugate momenta are defined as
\be
p_{\theta}=\frac{2 \dot \theta }{1+\kappa ^2 \cos ^2 \theta},\,\, p_{\psi}=\frac{\kappa\,  m \cos ^4 \theta \sin (2 \psi )+2 \cos ^2 \theta \dot \psi }{1+\kappa ^2 \cos ^4 \theta \sin ^2 \psi},
\ee
and we have identified the energy with, $E=-p_{t}=2 \, \dot t.$ We next find the Hamiltonian equation of motion using the ansatz mentioned in (\ref{ans1}). The phase space is defined by the four coordinates: $\{\theta, p_{\theta}, \psi,p _{\psi}\}.$ The constant energy surfaces ($E$)  are defined by the equation (\ref{vir2}). Keeping this in mind we solve the equations of motion, together with the special care that the Hamiltonian constraint is satisfied, for different values of $E$ and plot the  phase space trajectories for both the canonical pairs $\{\theta,p_{\theta}\}$ and $\{\psi,p_{\psi}\}.$ Surprisingly, we observe that for  generic initial conditions even in the presence of small $\kappa$ as we increase the energy the trajectories become chaotic. Initially, we identify that there is some kind of deformed tori in the phase space when energy is small but as we increase the energy these tori are completely destroyed and  the trajectories move freely in the phase space, the motion is only bounded by the total energy. We give the representative plots showing this behaviour below.  This further  supports our claim  stemming from the NVE analysis that even for small $\kappa$ the system shows some signature of chaos for $(\mathbb{R}\times S^5)_{\kappa}.$\par
\begin{figure}[!h]
		\centering
		\includegraphics[width=8.5cm]{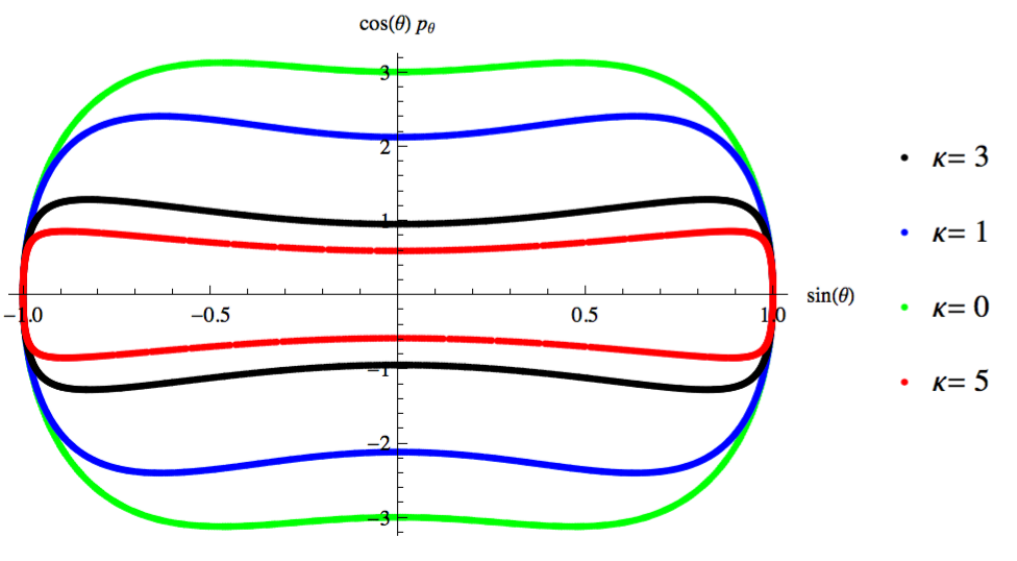}
		\caption{ The trajectories in $\{\theta,p_{\theta}\}$ plane  for $\{\psi(0)=0, p_{\psi}=0\}$ and $ m=2 , E=5.$}  \label{fig1}
\end{figure}

 
 First, for consistency check, we set the initial condition for  $\{\psi,p_{\psi}\}$ as  $\{\psi(0)=0,p_{\psi}(0)=0\}.$  So there will be no non-trivial phase space trajectories  in the $\{\psi,p_{\psi}\}$ plane, only we will have non-trivial trajectories in $\{\theta,p_{\theta}\}$ plane. In this case effectively what we are left with is a harmonic oscillator type system  characterized  by $\{\theta,p_{\theta}\}$  and we should not observe any chaotic behaviour for any values of $\kappa$ and $E$ (energy).  This has been displayed in Fig.~(\ref{fig1}), however here we plot $\{\sin\theta,\cos\theta~ p_{\theta}\}$ for better representation of the dynamics.  We can easily see that all the trajectories for different values of $\kappa$ are ordered.  One should mention, this is exactly what happens for motion in only $(\mathbb{R}\times S^3)_{\kappa}$ where we get exactly one harmonic oscillator with a $\kappa$ dependent mass and hence we observe no chaos whatever be the values of $\kappa$ and energy.

Next we choose more general boundary conditions where both the canonical pairs evolve. We plot the trajectories for both $\{\theta ,p_{\theta}\}$ and $\{\psi,p_{\psi}\}$ below for different values of $E$ and $\kappa$. In all cases, we have set the winding number $m=2$ for simplicity. 
\begin{figure}[!htb]
\centering
\begin{minipage}{.44\textwidth}
  \subsection*{$\kappa=0$}
  \centering
  \includegraphics[width=1\linewidth]{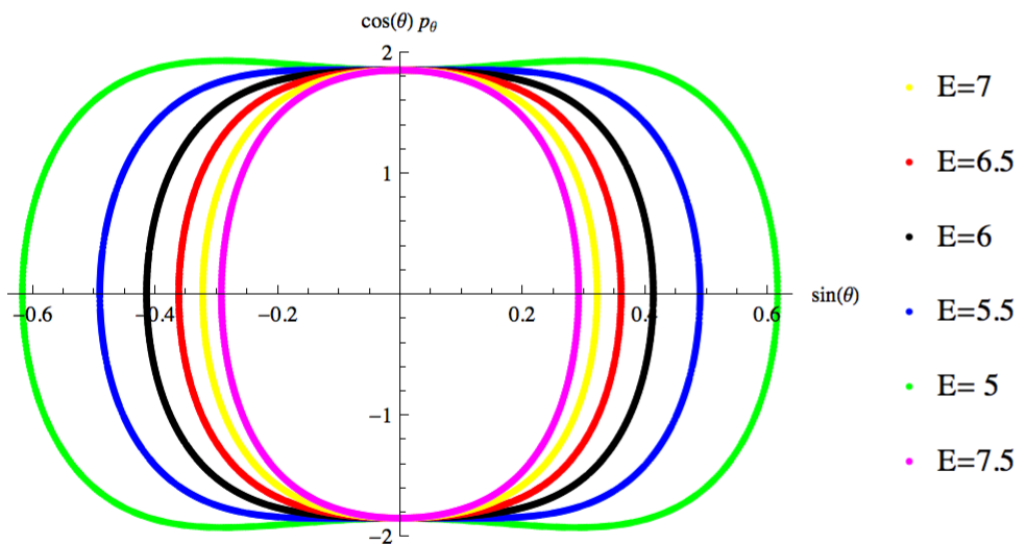}
\end{minipage}
\begin{minipage}{.44\textwidth}
  \centering
  \includegraphics[width=1.1\linewidth]{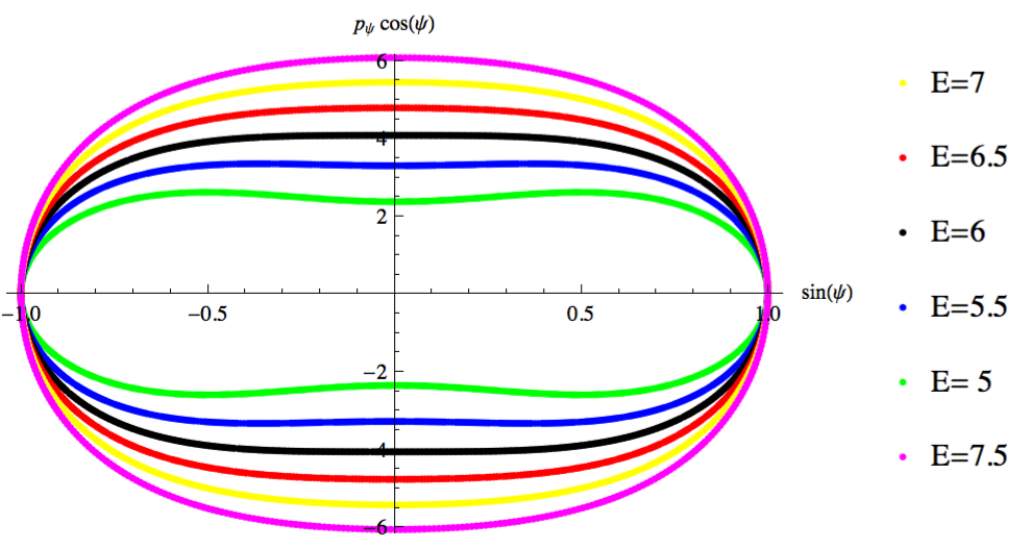}
\end{minipage}

\centering
\begin{minipage}{.44\textwidth}
 \subsection*{$\kappa=0.1$}
  \centering
  \includegraphics[width=1\linewidth]{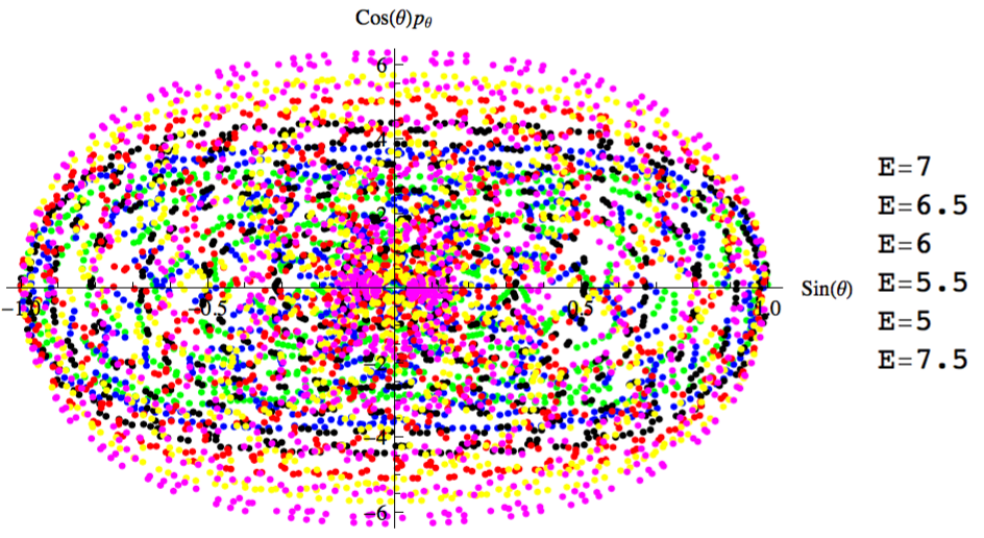}
\end{minipage}
\begin{minipage}{.440\textwidth}
  \centering
  \includegraphics[width=1.1\linewidth]{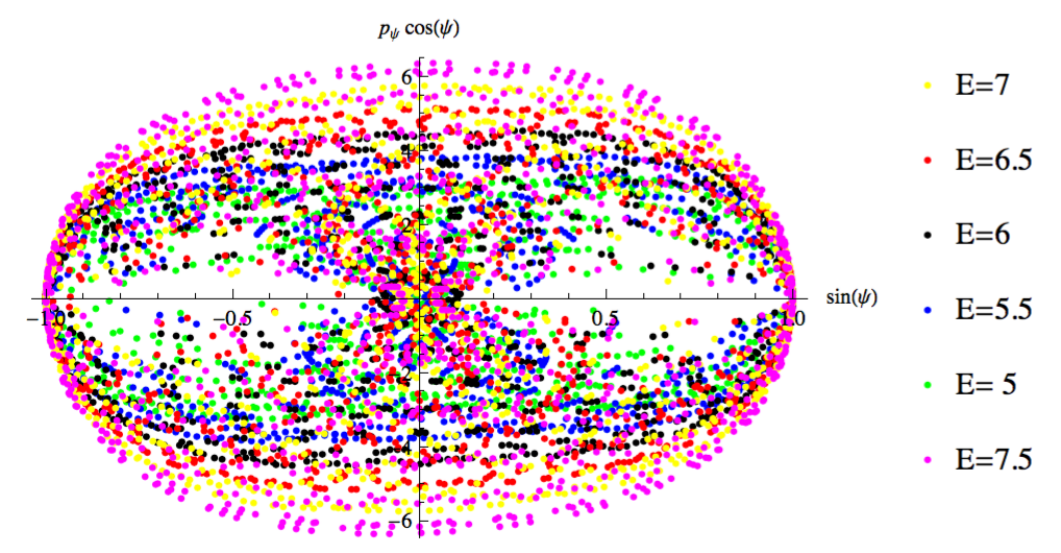}
\end{minipage}

\centering
\begin{minipage}{.44\textwidth}
 \subsection*{$\kappa=4$}

  \centering
  \includegraphics[width=1\linewidth]{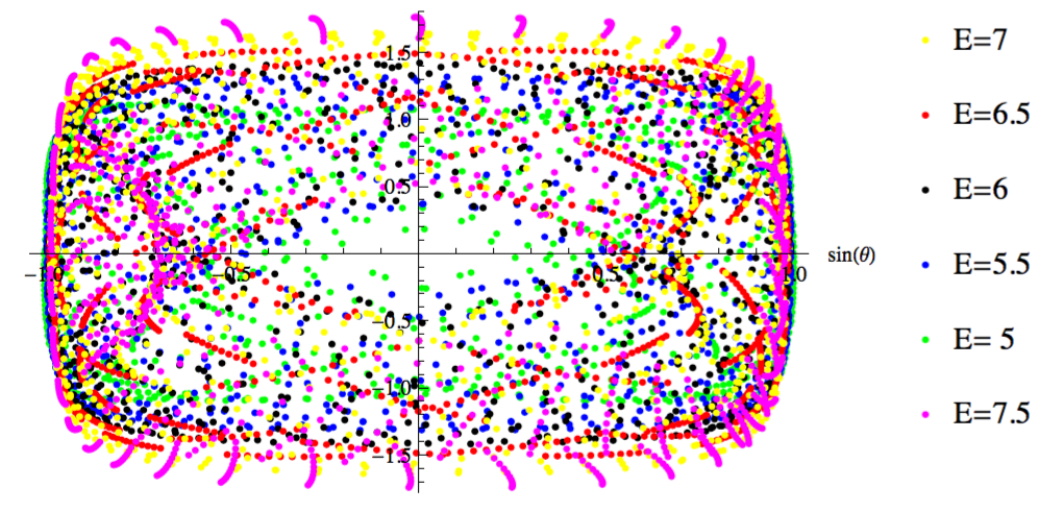}
\end{minipage}
\begin{minipage}{.44\textwidth}
  \centering
  \includegraphics[width=1.1\linewidth]{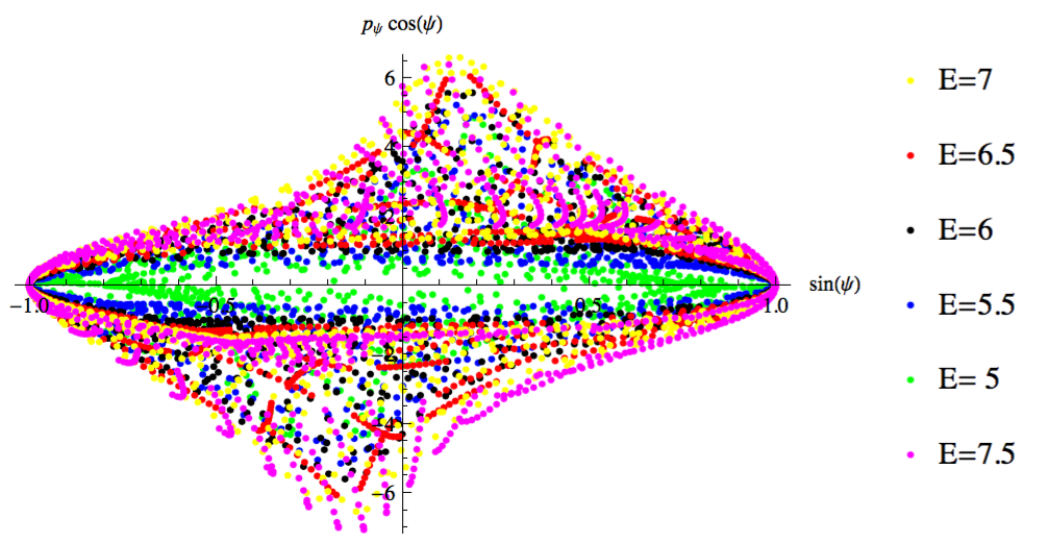}
\end{minipage}

\centering
\begin{minipage}{.440\textwidth}
\subsection*{$\kappa=10$}

  \centering
  \includegraphics[width=1\linewidth]{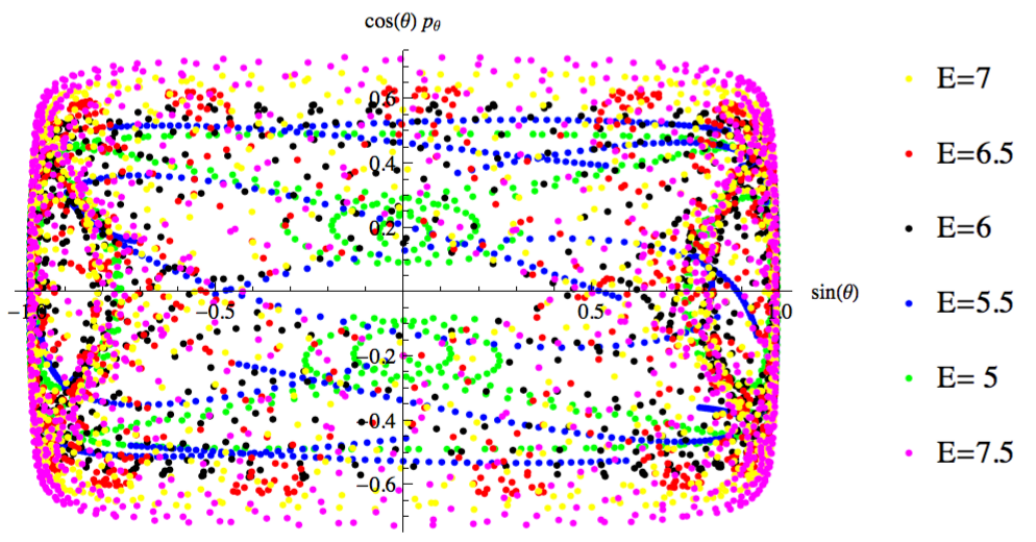}
\end{minipage}
\begin{minipage}{.440\textwidth}
  \centering
  \includegraphics[width=1.1\linewidth]{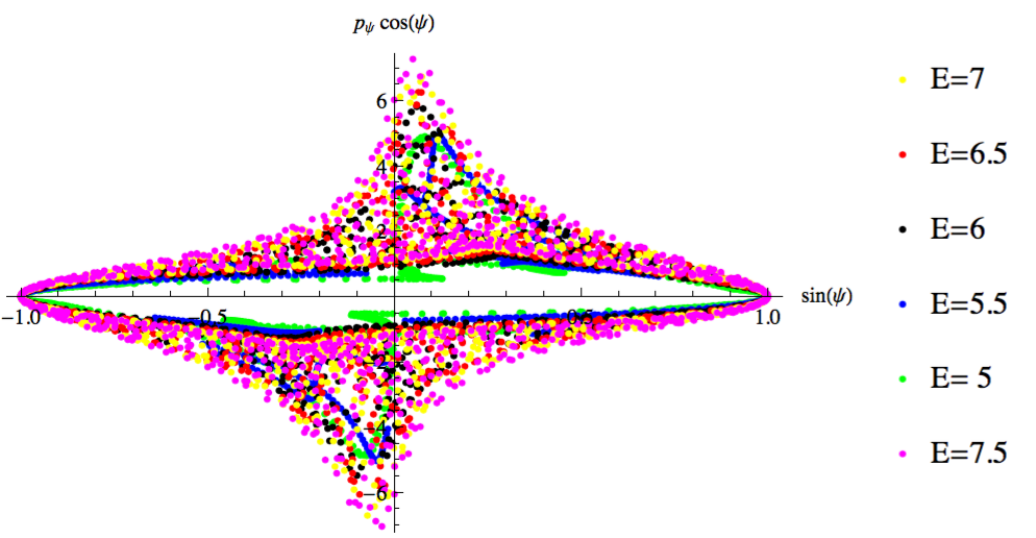}
\end{minipage}

\centering
\begin{minipage}{.440\textwidth}
\subsection*{$\kappa=100$}

  \centering
  \includegraphics[width=1\linewidth]{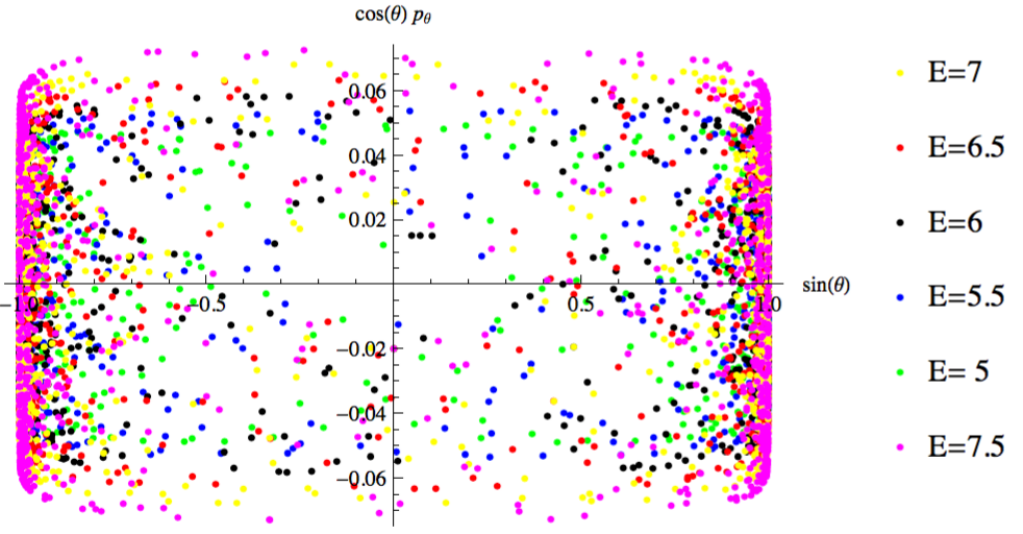}
\end{minipage}
\begin{minipage}{.440\textwidth}
  \centering
  \includegraphics[width=1.1\linewidth]{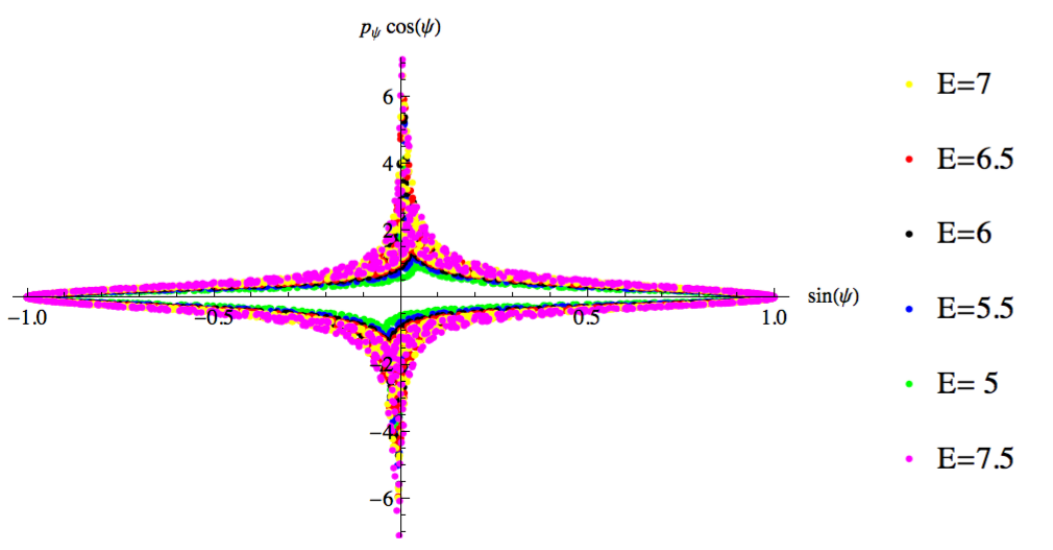}
\end{minipage}
\caption{ The phase-space trajectories for strings moving in the deformed five-sphere. Here, dynamics in the  $\{\theta,p_{\theta}\}$ and $\{\psi,p_{\psi}\}$ planes are plotted for various values of $\kappa$ and $E.$}
\label{fig2}
\end{figure}

All the plots in the left panel of  Fig.~(\ref{fig2}) show the phase space structure for $\{\theta,p_{\theta}\}$ and all the plots in the  right panel of  Fig.~(\ref{fig2}) show the phase space for $\{\psi,p_{\psi}\}.$ As it is evident from these plots that $\kappa=0$ trajectories are  ordered as expected, but even for small $\kappa, $ for example  if we consider $\kappa=0.1$ the trajectories become chaotic for all the values of the energy. The trajectories move freely in the phase space but within each of the energy envelopes (upto some numerical errors). Similarly we plot these phase space trajectories for higher values of $\kappa$ in subsequent plots. We see the chaos persists and for high values of $\kappa$, for example if we look at the $\kappa=100$ all the points in the phase space seem to concentrate near the edges of each of the energy contour. It is expected because the oscillators become very massive for higher values of $\kappa$ and the points in the phase space do not move much. This is also in agreement with our physical intuition. 
\section{Strings in the deformed $AdS$}
\subsection{Revisiting the case of $(AdS_3)_{\kappa}$}
We again revisit the case of strings in $(AdS_3)_{\kappa}$ as already discussed in \cite{Roychowdhury:2017vdo}, and start by putting $\zeta = \frac{\pi}{2}$ in the metric, which makes the NS-NS two-form zero. Moreover, identifying $\psi_2 = \psi$, we write the relevant metric in this case,
\be  \label{met}
ds^2_{(AdS_3)_{\kappa}} = \frac{1}{1-\kappa^2\sinh^2\rho}\left[ -\cosh^2\rho~ dt^2 + d\rho^2 \right]+\sinh^2\rho ~d\psi^2.
\ee 
Starting with a simple circular string ansatz as the following form,
\be
t= t(\tau) ,~~\rho = \rho(\tau),~~\psi = m\sigma,
\ee 
we can write down the equations of motion in the following form,
\bea
\cosh\rho \Big[2 \cosh \rho ~\ddot t \left(\kappa ^2-\kappa ^2 \cosh (2 \rho)+2\right)+8 \left(1+\kappa ^2\right)  \sinh \rho ~\dot t \dot\rho\Big] &=& 0, \nonumber \\
\sinh (2 \rho) \left[m^2 \left(1-\kappa ^2 \sinh ^2\rho\right)^2+\kappa ^2 \dot\rho^2+\left(1+\kappa ^2\right) \dot t^2\right]+2 \ddot\rho \left(1-\kappa ^2 \sinh ^2\rho\right) &=&0. \nonumber
\eea
These equations are trivially satisfied by $\rho = 0$ and $\dot\rho = 0$, which gives us an invariant plane to work with, provided we have a solution of the form,
\be
t(\tau) = \alpha\tau.
\ee
Also there is the Hamiltonian constraint to be satisfied,
\be
\frac{1}{1-\kappa^2\sinh^2\rho}\left[ -\cosh^2\rho~ \dot t^2 + \dot\rho^2 \right]+m^2\sinh^2\rho =0.
\ee
This in turn means that at $\rho = 0$, we should have $\dot t = 0$. Using the expansion
\be
\rho(\tau) =  0+\mathbf{r}(\tau),~~~ |\mathbf{r}|<<1,
\ee
 the desired NVE simply has the following form
\be
\ddot{\mathbf{r}} + \left[ m^2 +(1+\kappa^2)\dot t^2\right] \mathbf{r}= 0.
\ee
As we have discussed above, this is simply a Harmonic Oscillator equation of motion and hence is completely solvable.
\subsection*{A concrete example: Extended `Spiky' strings }
For the sake of completeness, we here mention the case of `spiky' strings \cite{Kruczenski:2004wg}  in $(AdS_3)_\kappa$ which, unlike circular strings, are extended object and has been well studied in the literature \cite{Banerjee:2015nha}. To discuss these strings in the Polyakov framework, the worldsheet embedding is quite involved, and has been discussed in \cite{Jevicki:2008mm}. We start here with that particular ansatz\footnote{Here we again note that the ansatz for such strings discussed in \cite{Roychowdhury:2017vdo}  i.e.
\be
t=\tau,~~\rho = \rho(\tau),~~\psi(\sigma,\tau)= m\sigma + \psi(\tau),
\ee
is seemingly incompatible with the Virasoro constraints. Notably, the constraint $T_{\tau\sigma} = 0$ leads to the condition
\be
\sinh^2\rho~m\dot\psi =0,
\ee
which forces either $m$ or $\dot\psi$ to be zero, rendering the ansatz inconsistent.
} of the form,
\be\label{spikeansatz}
t =\tau + f(\sigma),~~\rho = \rho(\sigma),~~\psi = \omega \tau + g(\sigma).
\ee
Note here the $AdS$ radial direction is not dependent on the worldsheet time coordinate. The equations of motion for $t$ and $\psi$ here gives rise to
\be
f'(\sigma) =\frac{C_1(1-\kappa^2\sinh^2\rho)}{\cosh^2\rho},~~~g'(\sigma) = \frac{C_2}{\sinh^2\rho}.
\ee
Where $C_{1,2}$ are just constants and primes denote derivative w.r.t $\sigma$. With this choice, the $\rho$ equation is given in turn by,
\bea
\frac{\sinh (2 \rho) \left(\left(1+\kappa ^2\right) \left(f'^2-1\right)+\kappa ^2 \rho'^2\right)}{\left(1-\kappa^2\sinh^2\rho\right)^2}+\sinh (2 \rho) \left(\omega^2-g'^2\right)+\frac{\rho ''(\sigma )}{1-\kappa^2\sinh^2\rho}=0. \nonumber
\eea
This is, as usual, supplemented by the Virasoro constraints. Here the constraint $T_{\tau\tau}+T_{\sigma\sigma}=0$ gives simply the Hamiltonian, which in turn is consistent with the $\rho$ equation. The other constraint $T_{\tau\sigma} =0$ gives a nice relation between the constants,
\be
-C_1+\omega C_2 =0.
\ee
In turn the constants can be chosen in the suggestive form,
\be
C_1=\frac{\omega}{2}\sinh 2\rho_1,~~C_2=\frac{1}{2}\sinh 2\rho_1,~~~\rho_1\neq 0.
\ee
Note that with our choice of constants, we can demand $\rho =\rho_1$ and $\rho' = 0$ is a solution for the above equation of motion. This $\rho_1$ is a constant depending on $\omega$ and $\kappa$. This is very significant in the construction since it exactly gives the point where the ``spike'' or the cusp occurs \cite{Jevicki:2008mm} on the string. For the deformed $AdS$, this constant was worked out in \cite{Banerjee:2015nha} and takes the value as following,
\be
\rho_1=\frac{1}{2}\cosh^{-1}\left[\frac{\left(-\kappa ^2+\omega ^2-1\right)-\sqrt{1-2 \left(2 \kappa ^2+1\right) \omega ^2+\omega ^4}}{\kappa ^2}\right]. 
\ee
Where taking the $\kappa\to 0$ limit properly, one could get back the position of the spikes in the case of undeformed $AdS$ i.e $\rho_1 = \coth^{-1}\omega$.
With these inputs and explicit expressions of $f'$ and $g'$ at hand, we can now expand the $\rho$ equation around the positions of the spikes as $\rho =\rho_1+ \mathbf{R}(\sigma)$, with $\mathbf{R}<<1$, to get the following NVE upto first order in $\mathbf{R}$,
\be
K_1 \mathbf{R}''+ K_2 \mathbf{R}+K_3 = 0.
\ee
The constant coefficients ($K_{1,2,3}$) are complicated combinations of $\omega, \kappa$ and functions of $\rho_1$, the total expressions of which are not really important for our purpose. This is certainly a solvable equation, and a representative Liouvillian solution can be given by, 
\be
\mathbf{R}(\sigma) = c_1 \sin \left(\frac{\sqrt{{K_2}} \sigma }{\sqrt{{K_1}}}\right)+c_2 \cos \left(\frac{\sqrt{{K_2}} \sigma }{\sqrt{{K_1}}}\right)-\frac{{K_3}}{{K_2}},
\ee
which determines the trajectory of the string along the $\sigma$ direction, with $c_{1,2}$ being constants. We note here that if we want to discuss dynamics of such extended string where the radial direction is time dependent, we can perform a  $\sigma \leftrightarrow \tau$ exchange in (\ref{spikeansatz}) to transform it to the `Dual Spike' solution \cite{Mosaffa:2007ty}, without any change in the analysis. 
\subsection{Analytical strings on $(AdS_5)_{\kappa}$}
 The most important exercise would be to study string motion in the full $(AdS_5)_{\kappa}$ space-time for reaching a concrete conclusion in our case.  To study circular strings in this background, we choose a particular simple ansatz as follows, 
\be \label{ans3}
t = t(\tau),~~\rho = \rho(\tau),~~\zeta = \zeta(\tau),~~\psi_1 =\psi_2 = m\sigma.
\ee
Note here we have put both winding numbers to be same for simplicity. With this choice, the Lagrangian for the system of strings take the expression,
\bea
\mathcal{L}_{AdS} &=& \frac{1}{1-\kappa^2\sinh^2\rho}\left[ -\cosh^2\rho~\dot{t}^2 + \dot\rho^2 \right] \nonumber
	\\
	&+& \frac{\sinh^2 \rho}{1+\kappa^2\sinh^4 \rho \sin^2\zeta}\left( \dot\zeta^2-m^2~\cos ^2\zeta \right) 
	-m^2~\sinh^2\rho \sin^2\zeta\nonumber
	\\
	&-&  \frac{\kappa\sinh^4\rho \sin (2\zeta)~m \dot\zeta}{1+\kappa^2 \sinh^4\rho \sin^2 \zeta} .
\eea\\
We will now write the explicit equations of motion for $\zeta$,
\bea
&&\sinh\rho \Big[\kappa ^2 m^2 \sinh ^5\rho \left(4 \kappa ^2 \sin ^5\zeta \cos\zeta \sinh ^4\rho -\sin (4 \zeta)\right)\\  \nonumber &&-8 \dot\rho \cosh \rho \left(\kappa  m \sin (2 \zeta) \sinh ^2\rho+\dot\zeta \left(\kappa ^2 \sin ^2\zeta \sinh ^4\rho-1\right)\right)\\ \nonumber && +4 \ddot\zeta  \sinh \rho \left(\kappa ^2 \sin ^2\zeta  \sinh ^4\rho+1\right)-2 \kappa ^2 \dot\zeta^2 \sin (2 \zeta) \sinh ^5\rho\Big]=0.
\eea
Also the equation for $\rho$ can be written as.
\bea
&&\frac{m^2 \cot ^2\zeta \sinh (2 \rho)}{\csc ^2\zeta+\kappa ^2 \sinh ^4\rho}-\frac{\kappa ^2 m^2 \sin ^2(2 \zeta) \sinh ^5(\rho) \cosh\rho}{\left(1+\kappa ^2 \sin ^2\zeta \sinh ^4\rho\right)^2}+m^2 \sin ^2\zeta \sinh (2 \rho)+\\ \nonumber && \frac{4 \kappa  m \dot\zeta \sin (2 \zeta) \sinh ^3\rho \cosh \rho}{\left(1+\kappa ^2 \sin ^2\zeta \sinh ^4\rho\right)^2}+\frac{\dot\zeta^2 \sinh (2 \rho) \left(\kappa ^2 \sin ^2\zeta \sinh ^4\rho-1\right)}{\left(1+\kappa ^2 \sin ^2\zeta \sinh ^4\rho\right)^2}+\frac{2 \ddot\rho}{1-\kappa ^2 \sinh ^2\rho}+\\ \nonumber &&\frac{\kappa ^2 \dot\rho^2 \sinh (2 \rho)}{\left(1-\kappa ^2 \sinh ^2\rho\right)^2}+\frac{\left(1+\kappa ^2\right) \sinh (2 \rho) \dot t^2}{\left(1-\kappa ^2 \sinh ^2\rho\right)^2} = 0.
\eea
Similarly, the $t$ equation has a form,
\bea
\cosh\rho \left(2 \cosh \rho \ddot t \left(\kappa ^2+\kappa ^2 (-\cosh (2 \rho))+2\right)+8 \left(1+\kappa ^2\right) \rho  \sinh \rho \dot t\right)=0.
\eea
These equations are not at all illuminating as one can see. Instead we can notice that the energy of this circular string is given by,
\be
E = \frac{\cosh^2\rho~\dot t}{1-\kappa^2\sinh^2\rho}.
\ee
We can show that the equations of motion in this case also vanish as in the previous one for $\rho=0$ and $\dot\rho = 0$, giving us an invariant plane to work with. Using this, we expand the $\rho$ equation of motion with,
\be
\rho(\tau) = 0+ \mathcal{R}(\tau),~~~~|\mathcal{R}|<<1.
\ee
Then, the $\rho$ NVE can be written in the following simple form upto the first order in $\mathcal{R}$,
\be \label{NVE2}
\ddot{\mathcal{R}}+\mathcal{R}\left(E^2 \left(\kappa ^2+1\right)+m^2-\dot\zeta^2\right) = 0.
\ee
This is a remarkably simple equation for such a complex  string background. Note that the effect of the singularity surface apparently vanishes here, since we are considerably closer to centre of the $AdS$ space. To find $\dot\zeta$ we notice that the conserved angular momentum associated to $\zeta$ can be written as
\be
J_{\zeta} = \frac{\sinh^2 \rho}{1+\kappa^2\sinh^4 \rho \sin^2\zeta}\dot\zeta - \frac{\kappa\sinh^4\rho \sin (2\zeta)~m }{2(1+\kappa^2 \sinh^4\rho \sin^2 \zeta)} .
 \ee
Near the invariant plane $\rho =0$, this can be shown to lead us to the expansion,
\be
\dot\zeta = \frac{2J_{\zeta}}{\mathcal{R}^2}+\mathcal{O}(\mathcal{R}^2).
\ee
As we have done before, we use the above in conjunction with (\ref{NVE2}) to write down the full form of the NVE. This equation, as evident, is completely solvable. A representative Liouvillian solution can be written as,
\be
\mathcal{R}(\tau)=\pm\frac{ \sqrt{4 J_{\zeta}^2 \left(E^2 \left(1+\kappa ^2\right)+m^2\right)-e^{-4 i \tau \sqrt{E^2 \left(1+\kappa ^2\right)+m^2}}}}{2 \sqrt{E^2 \left(1+\kappa ^2\right)+m^2} \sqrt{e^{-2 i \tau \sqrt{E^2 \left(1+\kappa ^2\right)+m^2}}}}.
\ee

 This analysis is strongly justified by the numerical calculations in the next section, where we show that no non-trivial dynamics appear for these $AdS$ strings in the phase space. Note that, throughout this section we have seen that the NVE for string motion in deformed $AdS$ spaces also become weakly deformed Harmonic Oscillator problems, indicating the inherent simplicity of the motion itself.

\subsection{Explicit numerical hamiltonian analysis }

We repeat the same analysis for this case as we have done in section (3.3) for the case of sphere. We will consider only the circular string profile as mentioned in (\ref{ans3}). We can leave the analysis for the extended spiky string for future investigation and instead focus only on these simple strings for our numerical experiment. The total Hamiltonian in this case  is, as expected, very complicated and we do not mention it here.
However, the two conjugate momenta are defined below,
\be
p_{\rho}=\frac{2 \dot \rho }{1-\kappa ^2 \sinh ^2 \rho },\,\,  p_{\zeta}= \frac{2 \dot \zeta  \sinh ^2 \rho -\kappa  \, m \sin (2 \zeta ) \sinh ^4 \rho}{1+\kappa ^2 \sin ^2 \zeta  \sinh ^4 \rho }
\ee
 and again the energy is identified as 
 \be
 E=p_{t}=\frac{2 \cosh ^2 \rho  \dot t }{1-\kappa ^2 \sinh ^2 \rho }.
 \ee
As before we solve the Hamiltonian equation of motions and we show the plots for the phase space trajectories for various values of $\kappa$ and energy $E$. Also from the metric (\ref{met}) we note that  there is a singularity surface at $\rho=\sinh ^{-1}\left(\frac{1}{\kappa }\right).$ So the larger the value of $\kappa$, the range of  $\rho$ becomes smaller and smaller as the trajectories never touch the singularity surface. The phase space plot of the evolution along $\rho$ will be restricted for each of the values of energy $E.$  Keeping this in mind we show the phase-space trajectories for both $\{\rho,p_{\rho}\}$ and $\{\zeta,p_{\zeta}\}$ plane. Also we set the winding number $m=2$ throughout our analysis. \begin{figure}[!htb]
\centering
\begin{minipage}{.45\textwidth}
  \subsection*{$\kappa=0$}
  \centering
  \includegraphics[width=1\linewidth]{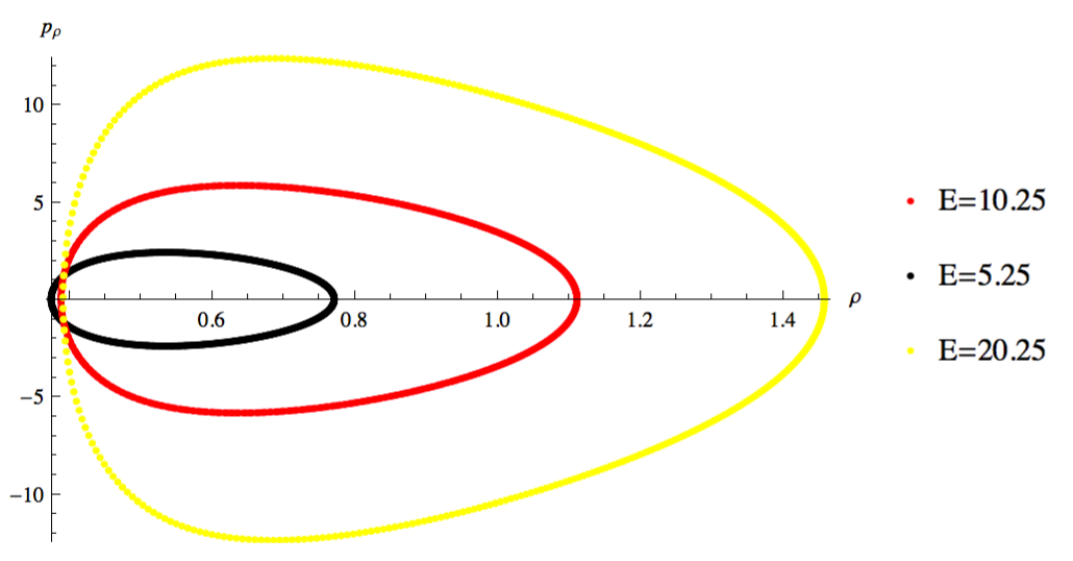}
\end{minipage}
\begin{minipage}{.45\textwidth}
  \centering
  \includegraphics[width=1.1\linewidth]{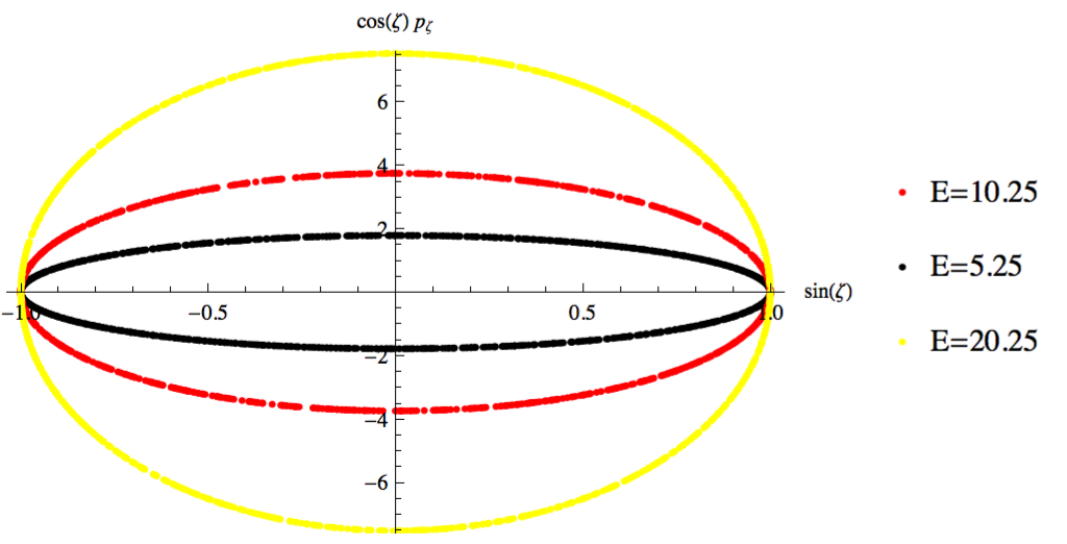}
\end{minipage}


\centering
\begin{minipage}{0.45\textwidth}
  \subsection*{$\kappa=1$}
  \centering
  \includegraphics[width=1.1\linewidth]{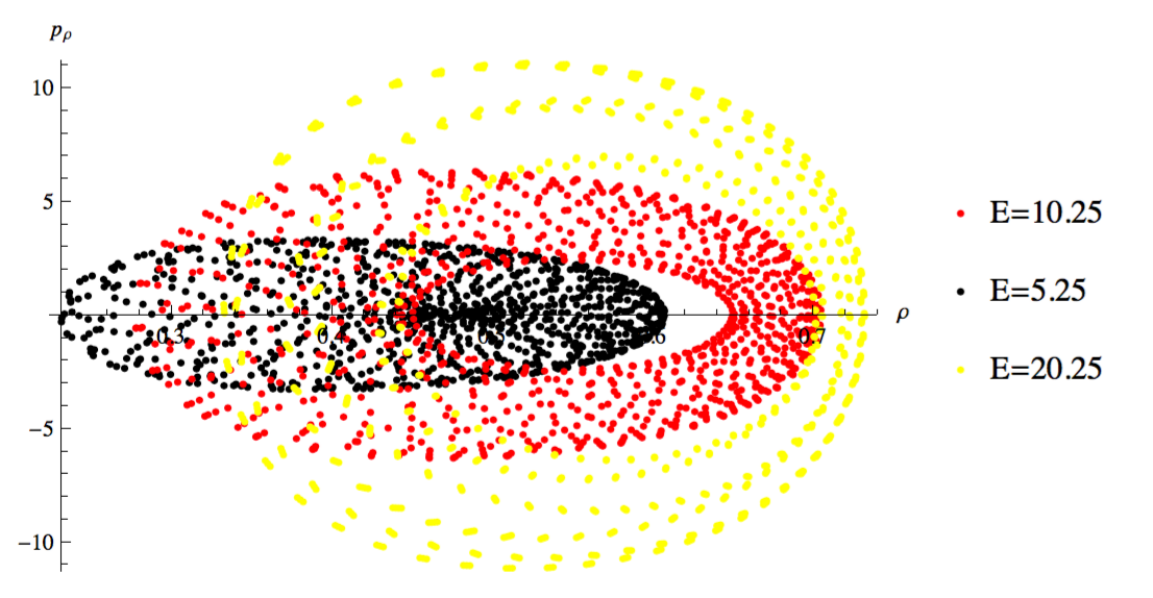}
\end{minipage}
\begin{minipage}{.45\textwidth}
  \centering
  \includegraphics[width=1.1\linewidth]{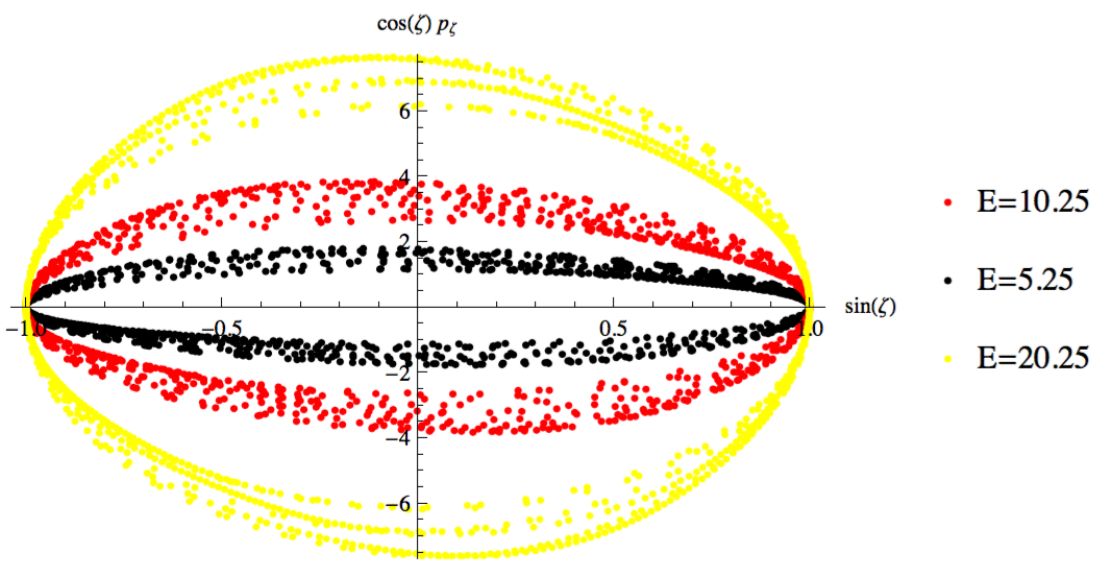}
\end{minipage}

\centering
\begin{minipage}{.44\textwidth}
  \subsection*{$\kappa=1.5$}
  \centering
  \includegraphics[width=1\linewidth]{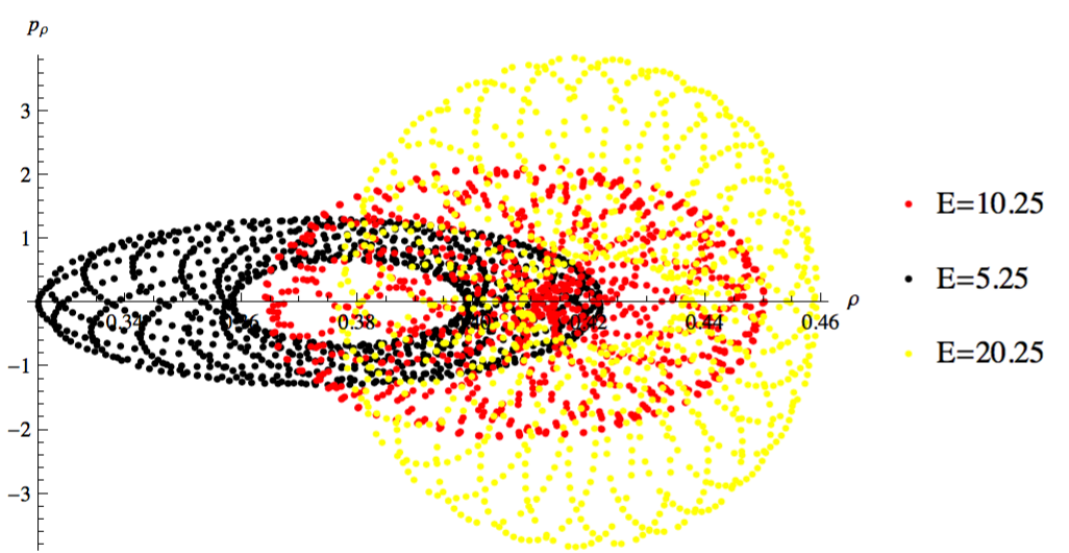}
  \label{fig:test1}
\end{minipage}
\begin{minipage}{.44\textwidth}
  \centering
  \includegraphics[width=1.1\linewidth]{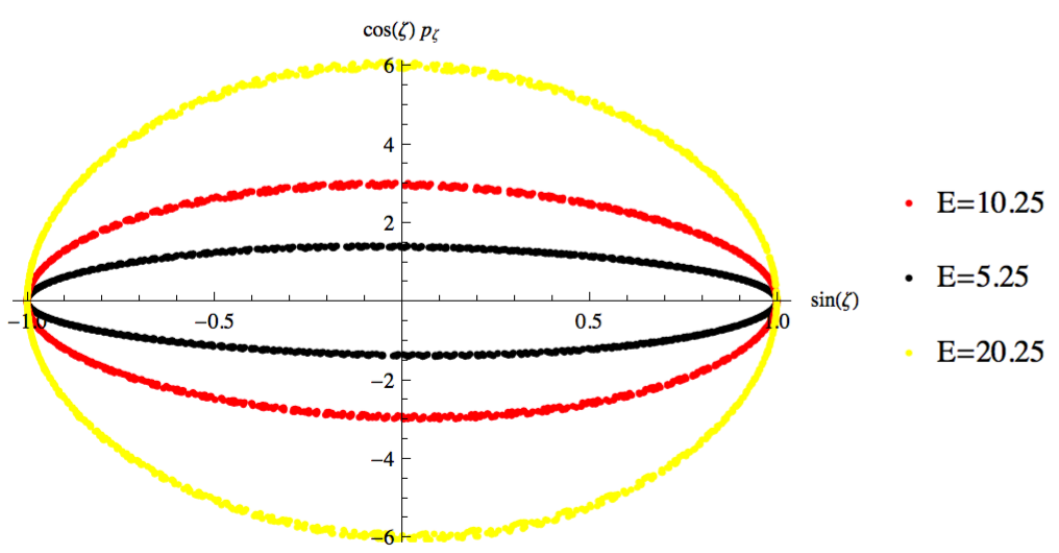}
\end{minipage}

\centering
\begin{minipage}{.44\textwidth}
  \subsection*{$\kappa=2$}
  \centering
  \includegraphics[width=1\linewidth]{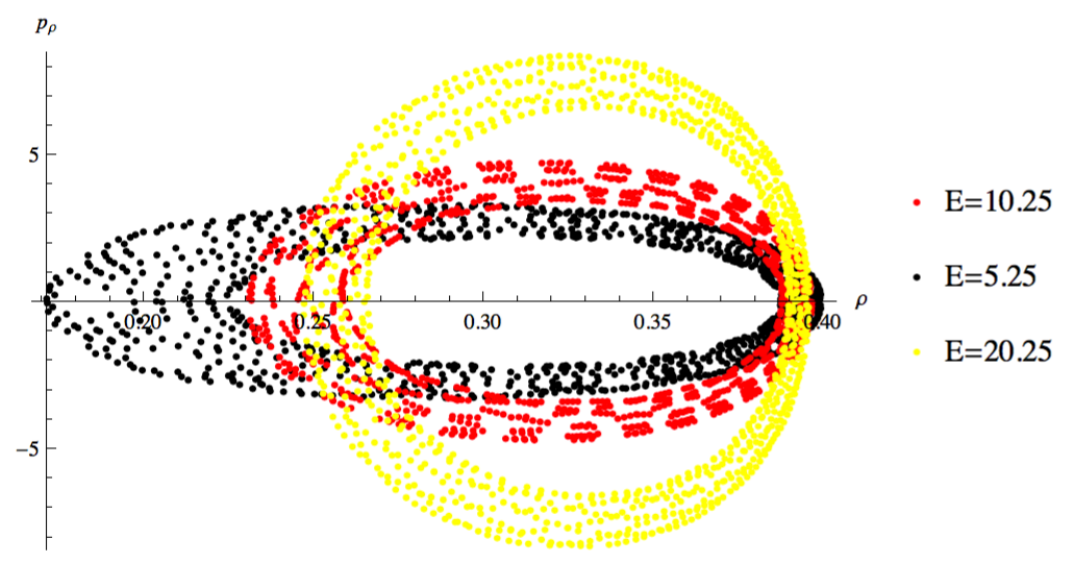}
  \label{fig:test1}
\end{minipage}
\begin{minipage}{.44\textwidth}
  \centering
  \includegraphics[width=1.1\linewidth]{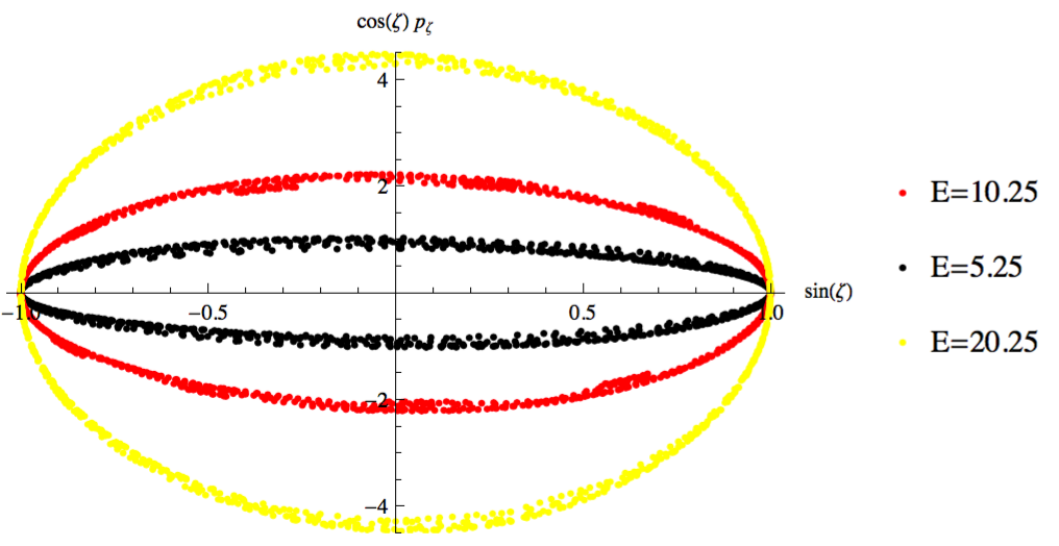}
\end{minipage}
\caption{ These plots display  the phase-space trajectories for  $\{\rho,p_{\rho}\}$ and $\{\zeta,p_{\zeta}\}$ plane for various values of $\kappa$ and $E.$}
\label{fig3}
\end{figure}

As expected for $\kappa=0$ the trajectories are ordered. Also from the Fig.~(\ref{fig3}) we can easily infer that for AdS within the range of the values of $\kappa$ and $E$ that we have considered the trajectories remains always ordered (upto some numerical errors) and hence it is in good agreement with our conclusion from NVE analysis that unlike the $S^5_\kappa$ case, there is no chaotic behaviour for the case of $AdS_\kappa$. 
\section{Summary and Conclusion}
Let us first summarize the paper briefly. In this note, we set out on a humble quest, to settle the issue of analytic/numeric integrability techniques clashing with the well-known algebraic formulation for Yang-Baxter deformed $AdS_5\times S^5$ case. Although the methods of studying classical integrability of an exact string background have been believed to be equivalent to each other, we find several pieces of evidence suggesting the contrary. Starting from revisiting the calculations provided in \cite{Roychowdhury:2017vdo}, we conclude there is no such analytical/numerical evidence of chaotic motion appearing in the string phase space for $(AdS_3)_\kappa$ and $(\mathbb{R}\times S^3)_\kappa$ case. However, we find that for a rigidly spinning circular string moving in  $(\mathbb{R}\times S^5)_\kappa$, the motion surprisingly runs into chaos when we turn on a non-zero value of $\kappa$. We show both analytically via perturbations around the classical trajectories and via numerical experiments that irregular evolution of trajectories indeed occurs in this case. Surprisingly, the case of $(AdS_5)_\kappa$, which has a prominent space-time singularity, doesn't show any evidence of chaotic string motion as the trajectories never reach the singularity itself. 

This rather shocking revelation puts us in crossroads about how we view the notion of classical integrability in this case from different vantage points. If there exists one such dynamical model truncation for the string system, where the differential equations are not integrable, the phase space definitely has problems. Non-integrability often does not explicitly lead to chaos, but in our case, it is evident in the Poincare sections. There could be some added subtlety to the case of $(\mathbb{R}\times S^5)_\kappa$, which is not captured by our analysis here, and which could stabilize the solution against irregular perturbations. However, the idea of what that could be is eluding us as of now. Another viable point that can be considered, comes from the discussion presented in \cite{Banerjee17}. There, it was explicitly showed that at the fast spinning string limit the equations of motion for strings in $(\mathbb{R}\times S^5)_\kappa$ maps to that of a complex $\beta$-deformed sphere. This background has been shown to be classically non-integrable via analytical/numerical techniques \cite{Giataganas2013}. We speculate that this might have deeper implications that we have had thought earlier, although in this paper we are not exactly taking the fast spinning limit anywhere.


We reiterate that the results presented in this paper are in no way a conclusive ``proof'' of non-integrable dynamics in the deformed background, the integrability of which has been proven beyond doubt in series of investigations via algebraic methods. However, the string motion in deformed five-sphere concerned here are consistently constructed and they indeed seem to show chaotic behaviour at the level of differential equations.  Hence, the general question about disparity of different methods to check classical (non)-integrability, however, persists strongly. The implication of our findings is that this disparity $does$ exist. This work is standalone example towards scratching the surface of this mystery, which surely needs further attention. One might try to find an answer to this via exploring other well-known but non-trivial classical string backgrounds. A very useful exercise perhaps would be to study the BTZ black hole background. BTZ has been known to be classically integrable \cite{David1} for few years now. But since this background contains event horizons, one would easily guess that string motion becomes irregular near these horizons. The classical string solutions of these type can also be circular provided they
wind around the horizon and then eventually fall into it \cite{David:2010yg}. One could then investigate such string motion in BTZ background using the procedures used in this paper. This might give more insight into this apparent disparity of discussing string integrability in this context. However, all of this still remains speculations, and certainly require rigorous understanding. We plan to come back to these concerns in the near future. 
 
\section*{Acknowledgements}
The authors would like to thank Carlos Nunez, Arkady Tseytlin and Oleg Lunin for useful comments. We also thank Dimitrios Giataganas for useful correspondence and suggestions.  Aritra Banerjee (ArB) is supported in part by the Chinese Academy of Sciences (CAS) Hundred-Talent Program, by the Key Research Program of Frontier Sciences, CAS, and by Project 11647601 supported by NSFC. ArB would like to thank Universite Libre Bruxelles (ULB) and CERN theoretical physics for kind hospitality during the final stages of this work. AB is supported by JSPS fellowship and  JSPS Grant-in-Aid for JSPS fellows 17F17023. AB would like to thank Physics Department, University of Murcia, IFT Madrid, GGI Florence, Physics Department, University of Geneva and Max Planck Institute for Gravitational Physics, Golm, Germany  for kind hospitality during the final stages of this work.
\section*{Appendix: Effect of $\kappa\neq 0$ in NVE via Kovacic algorithm }
We here study the NVE  for $\theta (\tau)$ for the case of $(R\times S^{5})_{\kappa}$ analytically. We will apply Kovacic's algorithm to study in detail whether it will have a Liouvillian solution or not. We start with the $\theta$ equation of motion as written in (\ref{theq}). As we have seen earlier, $\theta =0, \dot\theta =0$ is a trivial solution for this equation of motion. So, to find the NVE we would consider fluctuation around this solution. At this point, the Virasoro constraint gives us the equation of $\psi$ on this invariant plane,
\be
\dot\psi^2 = E^2(1+\kappa^2\sin^2\psi)-m^2\cos^2\psi.
\ee
The solution for $\psi(\tau)$ on this invariant plane can then be written as a Jacobi function,
\be
 \cos\psi(\tau) = \mathbf{sn}~\Big[ \sqrt{1+\kappa^2}E\tau~|~\frac{E^2\kappa^2+m^2}{E^2(1+\kappa^2)}\Big].
\ee
 Then expanding the equation of motion up to linear order in the fluctuation $\theta(\tau)= 0+y(\tau)$, we obtain the required NVE,
 \be
\frac{2 \ddot y(\tau) }{1+\kappa ^2}+y(\tau) \left[\frac{m^2 \left(\kappa ^2 \sin ^2(2 \psi)-4 \cos ^2\psi\right)+8 \kappa  m \dot\psi \sin (2 \psi)+4\dot\psi^2 \left(1- \kappa ^2 \sin ^2\psi\right)}{2 \left(1+\kappa ^2 \sin ^2\psi\right)^2}\right]=0. \ee

 Now we perform the change of variable as $\tau \to z=\cos \psi(\tau)$ to algebrize the NVE and rewrite it as a differential equation with rational coefficients. After some involved algebra, the  above equation takes the following form.
\begin{align}
\begin{split}\label{diff1}
y''(z)+\mathcal{B}(z) y'(z)+\mathcal{A}(z) y(z)=0,
\end{split}
\end{align}
where the primes denote derivatives w.r.t $z$ and the coefficient functions are,
\begin{align}
\begin{split}
&\mathcal{B}=z \left[\frac{E^2 \kappa ^2+m^2}{z^2 \left(E^2 \kappa ^2+m^2\right)-E^2 \left(1+\kappa ^2\right)}+\frac{1}{z^2-1}\right],~
\mathcal{A}=\frac{\mathcal{C}}{\mathcal{D}};\\&
\mathcal{C}=\left(1+\kappa ^2\right) \Big[-16 \, z\,\kappa \, m \left(1-z^2\right) \left(E^2 \left(1+\kappa ^2\right)-z^2 \left(E^2 \kappa ^2+m^2\right)\right)\\&+\left(1-z^2\right) \left(4 \kappa ^2 \left(z^2-1\right)+4\right) \left(E^2 \left(1+\kappa ^2\right)-z^2 \left(E^2 \kappa ^2+m^2\right)\right)^2\\&+4 m^2 z^2 \left(\kappa^2-\kappa^2 z^2-1\right)\Big],\\&
\mathcal{D}=4 \left(z^2-1\right) \left(\kappa ^2 \left(z^2-1\right)-1\right)^2 \left[(E^2\kappa^2+m^2)z^2-E^2 \left(1+\kappa^2\right)\right].
\end{split}
\end{align}
Now we can see that  $\mathcal{A}$ and $\mathcal{B}$ are rational polynomial of $z$. Then we can bring the equation in the following normal or Schrodinger form
\be\label{diff2}
w''(z) + V(z)w(z) = 0,
\ee
via the transformation of variables as,
\be
y(z)\rightarrow w(z) = y(z)\text{exp}\left[-\frac{1}{2}\int^z \mathcal{B}(x)~dx  \right].
\ee
 The potential for this equation can be given by,
\be
V(z)=-\frac{1}{4}\Big(2\mathcal{B}'+\mathcal{B}^2-4\mathcal{A}\Big).
\ee
Explicitly,
\begin{align}
\begin{split}
V(z)&=-E^2 \kappa ^2+\frac{3 z^2 \left(E^2-m^2\right)^2}{4 \left(z^2-1\right)^2 \left(E^2 \left(\kappa ^2 \left(z^2-1\right)-1\right)+m^2 z^2\right)^2}-\frac{2 z^2 \left(m z \left(z^2-1\right) (2 \kappa +m z)+1\right)}{\left(\kappa ^2 \left(z^2-1\right)^2-z^2+1\right)^2}\\&-\frac{(E^2-m^2) \left(4 E^2-m^2 \left(3 z^2+1\right)\right)}{2 m^2 \left(z^2-1\right)^2 \left(E^2 \left(\kappa ^2 \left(z^2-1\right)-1\right)+m^2 z^2\right)}+\frac{z^4 \left(-\left(E^2+m^2\right)\right)+E^2+m^2 z^2+1}{\left(z^2-1\right)^2}\\&+\frac{E^2 \left(2-2 m^2 z^2 \left(z^2-1\right)\right)+m^3 (-(z-1)) z (z+1) \left(4 \kappa +m z \left(z^2+2\right)\right)-m^2 \left(z^2+2\right)}{m^2 \left(z^2-1\right)^2 \left(\kappa ^2 \left(z^2-1\right)-1\right)}.
\end{split}
\end{align}
Now (\ref{diff2}) is a linear second order differential equation and also in the correct form for applying Kovacic algorithm \cite{Kovacic}  to test whether it admits Liouvillian solutions. Now according to Kovacic algorithm, the potential should  at least satisfy one of the following three necessary (but not sufficient) criteria so that  the differential equation (\ref{diff2}) and hence (\ref{diff1})  will admit Liouvillian solution.
\begin{itemize}
\item{I: All the poles of $V(z)$ will be either of order 1 or  even order and the order of $V(z)$ at infinity has to be either even or gerater than 2.  The order of $V(z)$ at infinity can computed by the subtraction of the highest power of $z$ in numerator from the highest power of $z$ in denominator. }
\item{II: All the poles of  $V(z)$ will be of odd order greater than 2 or it will posses just one pole of order 2. \footnote{We thank Carlos Nunez for explaining this point to us.} }
\item{III: Order of all the poles of $V(z)$ are less than or equal to $2$ and order of $V(z)$ at infinity has to be at least order 2.}
\end{itemize} 

These conditions can be proven to be equivalent to the differential-Galois group treatment for differential equations. Now we can check that the potential $V(z)$ mentioned in (\ref{diff2}) violates all of these three conditions and hence our normal variation equation for the $\theta$ does not admit  Liouvillian solutions. This is consistent with our numerical results  presented in the section (3.2). Putting $\kappa = 0$ one can easily check that this process succeeds as one would expect for the undeformed five sphere.

\end{document}